\documentclass[12pt]{article}

\usepackage{newtxtext,newtxmath}
\usepackage{hyperref}
\usepackage{multirow}     % often needed in reports
\usepackage{graphicx}
\usepackage[letterpaper,margin=1in]{geometry}

\renewenvironment{abstract}
	{\quotation}
	{\endquotation}

\date{}

\makeatletter
\renewcommand{\fnum@figure}{\textbf{Figure \thefigure}}
\renewcommand{\fnum@table}{\textbf{Table \thetable}}
\makeatother

\usepackage{scicite}

\usepackage{url}

\def\scititle{Personalized Large Language Models Can Increase the Belief Accuracy of Social Networks
}

\title{\bfseries \boldmath \scititle}

\author{
	Adiba Proma$^{1\ast \dagger}$,
	Neeley Pate$^{1\dagger}$,
        Sean Kelty $^{2}$, 
        \and
        Gourab Ghoshal$^{2}$,
	James N. Druckman$^{3}$, 
	Ehsan Hoque$^{1}$ \and
	\small$^{1}$Department of Computer Science, University of Rochester, Rochester, 14627, USA. \and
	\small$^{2}$ Department of Physics and Astronomy, University of Rochester, Rochester, 14627, USA. \and
        \small$^{3}$ Department of Political Science, University of Rochester, Rochester, 14627, USA. \and
	\small$^\ast$ Adiba Mahbub Proma. Email: aproma@cs.rochester.edu \and
	\small$^\dagger$These authors contributed equally to this work.
}

\begin{document} 

% % Insert the title and author list
\maketitle

% Abstract, in bold
% There are strict length limits, and not all formats have abstracts.
% Consult the journal instructions to authors for details.
% Do not cite any references in the abstract.
\begin{abstract} \bfseries \boldmath

Large language models (LLMs) are increasingly involved in shaping public understanding on contested issues. This has led to substantial discussion about the potential of LLMs to reinforce or correct misperceptions. While existing literature documents the impact of LLMs on individuals’ beliefs, limited work explores how LLMs affect social networks. We address this gap with a pre-registered experiment (N = 1265) around the 2024 U.S. presidential election, where we empirically explore the impact of personalized LLMs on belief accuracy in the context of social networks. The LLMs are constructed to be personalized, offering messages tailored to individuals’ profiles, and to have guardrails for accurate information retrieval. We find that the presence of a personalized LLM leads individuals to update their beliefs towards the truth. More importantly, individuals with a personalized LLM in their social network not only choose to follow it (as a way of indicating they would like to obtain information from it in subsequent interactions), but also construct subsequent social networks to include other individuals with beliefs similar to the LLM – in this case, more accurate beliefs. Therefore, our results show that LLMs have the capacity to influence individual beliefs and the social networks in which people exist, and highlight the potential of LLMs to act as corrective agents in online environments. Our findings can inform future strategies for responsible AI-mediated communication.

\end{abstract}

\noindent

Misinformed beliefs pose major challenges across a range of domains, from public health and civic trust to the understanding of basic facts \cite{nyhan2020facts, matatov2022stop, oehmichen2019not, bovet2019influence}. These beliefs can erode confidence in voting systems \cite{lazer2018science, tucker2018social}, vaccination programs \cite{Eggertson_2010, Skafle_NordahHansen_Quintana_Wynn_Gabarron_2022}, and mainstream news organizations \cite{Ognyanova_Lazer_Robertson_Wilson_2020}. They also are often tied to misperceptions about policies and empirical reality \cite{nyhan2020facts}.

The rise of large language models (LLMs) has amplified both the risks and opportunities in this space. LLMs can generate misinformation at scale \cite{huang2024fakegpt}, sometimes producing false claims that humans find difficult to distinguish from truth \cite{chen2024can}. At the same time, LLMs are being explored for their potential to support factual accuracy through automated fact verification \cite{pan2023fact, wang2024mmidr}, disinformation detection \cite{lucasetal2023fighting}, and conversational engagement \cite{costello2024durably}.

Central to both their promise and their risk is the persuasive nature of LLMs. Empirical work has shown that LLMs can be highly persuasive on their own \cite{breum2024persuasive}, sometimes rivaling trusted human authorities such as public health agencies \cite{karinshak2023working}. LLMs can also be personalized --- that is, LLMs can structure communications to match a person’s reasoning styles \cite{costello2024durably} or psychological profiles \cite{matz2024potential}. Such personalized LLM 
messages are often more effective than general non-personalized ones \cite{matz2024potential, simchon2024persuasive}, especially if they match their audience's personality \cite{teeny2024we}, and can exert more influence than human-generated messages \cite{matz2024potential}. A crucial limitation of existing work on LLMs and belief formation is its exclusive focus on one-on-one interactions, where a user directly engages with the model in isolation \cite{costello2024durably}.
%Yet, virtually all work focuses on one-on-one interactions, where a user directly engages with the model in isolation \cite{costello2024durably}.

In reality, beliefs are rarely formed or revised in isolation. They are shaped by social context with exposure to multiple information streams \cite{PENNYCOOK2021388, Bolsen_Druckman_Cook_2014, Ahmed_Tan_2022}. Prior work has shown that social networks and peer influence play critical roles in belief formation and resistance to correction \cite{matatov2022stop, bovet2019influence, chen2021persuasion, fazio2024combating}. Despite this, little is known about how personalized LLM-generated content might function in networked environments where users are exposed to diverse and competing signals. Moreover, users actively construct their social networks, and how exposure to a personalized LLM might influence that process has gone unstudied.

Here, we study the impact of personalized LLMs in the context of a social network. The LLMs are constructed to provide accurate information, in line with work on misperception corrections \cite{porter2024factual}. Specifically, we expect that, relative to individuals in social networks without a personalized LLM, those in networks with access to a personalized LLM response will be more likely to move toward belief accuracy (hypothesis 1). This follows since the LLM will provide persuasive messaging, tailored to individual preferences. Additionally, we predict that those in social networks with an LLM are likely to include the LLM in their subsequent social network at rates greater than chance (hypothesis 2a). Prior research suggests that individuals tend to seek factual truth  \cite{pennycook2021shifting, sinclair2012social}, and LLMs, equipped with credible information, might be better able to facilitate that shift, making them desirable agents for  subsequent engagement \cite{}. Finally, this same logic (i.e., the truth-seeking nature of individuals) leads us to hypothesize that, relative to individuals in social networks without an LLM (i.e., the control), those with access to an LLM will form subsequent social networks that include significantly more members whose beliefs tend towards accuracy (hypothesis 2b). The belief updating a la hypothesis 1 combined with the desire to associate with others similar to themselves contributes to this prediction. This prediction suggests that adding an LLM to a network not only stimulates belief accuracy but it also shapes subsequent social networks such that they include more accurate beliefs overall.

\begin{figure}[h] % Do not use \begin{figure*}
	\centering
	\includegraphics[width=0.95\textwidth]{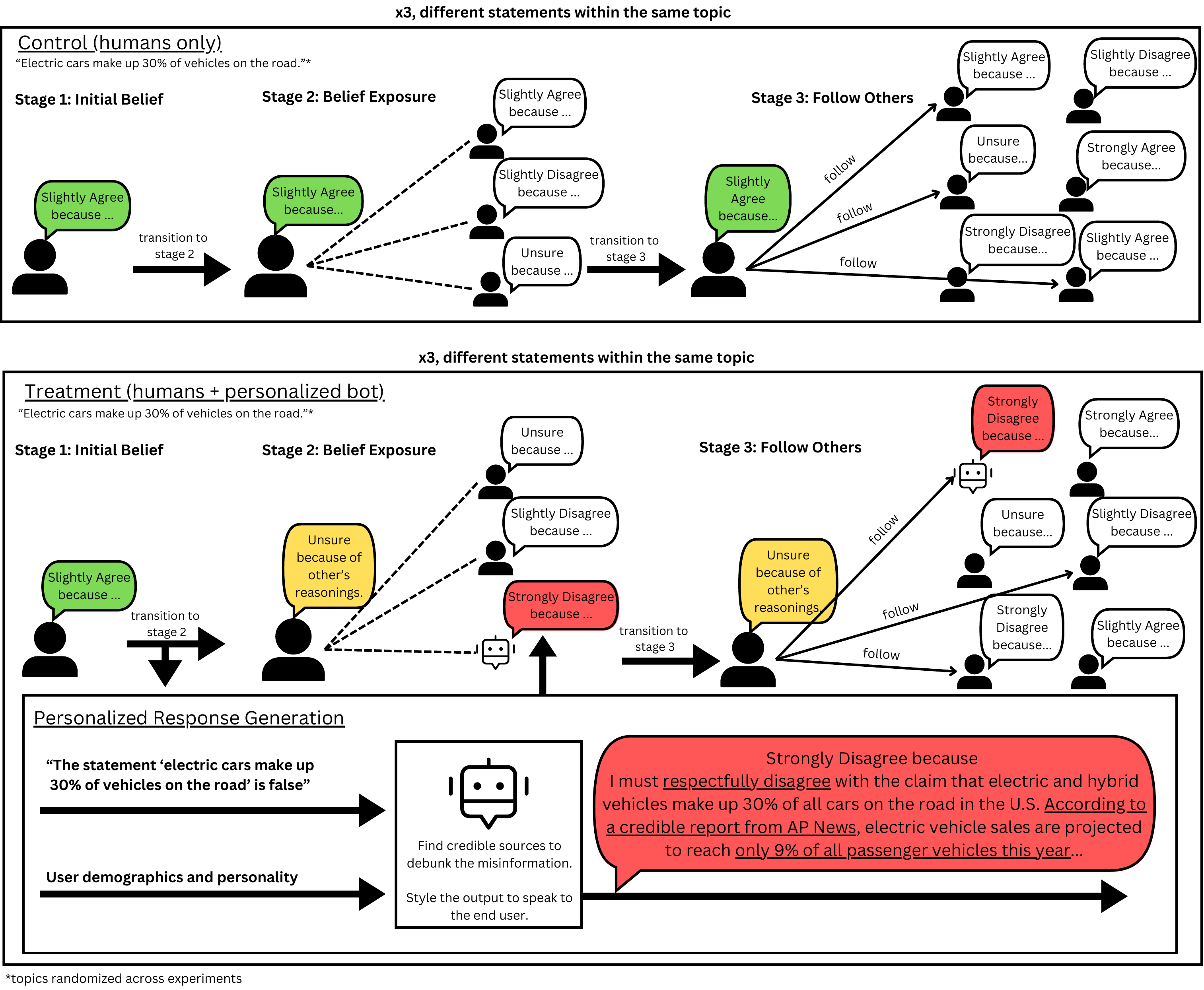} % for an image file named example_figure.*
	% Pick an appriopriate width for the size of the image

	% Captions go below figures
	\caption{\textbf{Schematic of the Experimental Setup.}
		The schematic shows the control and the treatment conditions. The control has only humans and the treatment has a personalized bot for each human participant. \textbf{TOP PANEL: Control.} In the control, the individual first encounters a statement and provides a rating based on believability. Then, they are exposed to others' views and given the opportunity to update their belief. Finally, they are shown additional individuals and given the opportunity to follow or unfollow them. This repeats three times. \textbf{BOTTOM PANEL: Treatment.} The same setup is used for the treatment, where each person within the study has a personalized bot assigned to them. The bot is given the prompt statement, the factuality, and the demographics and personality of the individual. It then generates a tailored response supported by news articles and rhetorical styling. The user is then shown the bot response along with the responses of others. The experiments used 4 different topics: overpopulation, immigration, fuel and oil, and inflation. Each experiment used 3 different statements, with 2 being false and 1 being true.}
	\label{fig:ExpSchematic} % give each figure a logical label name
\end{figure}

Here, we evaluate our hypotheses in a dynamic, socially embedded, empirical setting (Figure~\ref{fig:ExpSchematic}). We combine three elements rarely studied together: real-time belief updated, network-mediated exposure, and personalized LLM messaging. Specifically, we were interested in whether the presence of the LLM caused participants to have more accurate beliefs, and if it impacted how participants constructed their social networks. 

In a pre-registered experiment (N = 1265) conducted from October to December 2024 (see Supplemental Materials (SM), Section \ref{part_demographics} and Tables~\ref{tab:demographics}, \ref{tab:chi2_demographics}, and \ref{tab:mwu_personality}), individuals rated on a five-point scale the extent to which they believed a randomly assigned political statement that could be objectively true or false (see SM Subsection \ref{PoliticalStatement}, and Tables~\ref{statements_part1}, \ref{statements_part2}, \ref{statements_part3}). They also provided a rationale for their belief (see SM Figure \ref{fig:stage1} for examples). Next, participants were exposed to other users' responses and had an opportunity to update their beliefs. In the control condition, participants only saw three responses from peers who had simultaneously rated the same statement. In the treatment condition, participants saw two such peer responses, and an additional personalized response from the LLM bot. Finally, in the third stage, participants were shown six overall responses. These include the responses of those they previously followed, the bot in case of treatment, and the rest randomly `recommended' by the system. Participants were instructed to follow overall three out of these six entries, and could subsequently choose to follow or unfollow others (including the bot), allowing us to observe changes in the network structure (see \textit{Materials and Methods},  and Supplemental Figures \ref{fig:stage1}, \ref{fig:stage2}, \ref{fig:stage3}). Participants were told they may or may not interact with an LLM bot in the experiment, and had provided consent to this setup. All experiments were hosted in real time with 10-28 participants, based on participant turnout, and each experiment had three rounds.

Off-the-shelf LLMs, while shown to be persuasive \cite{breum2024persuasive}, are not personalized. Therefore, we design the personalized LLM bot such that given a statement and its factuality, it can personalize the argument considering individual’s preferred news sources, and rhetorical style (see SM, Section \ref{exp_llm_design_supp}, \ref{tablearticles}, \ref{tab:llmsamples1}, \ref{tab:llmsamples2}, \ref{tab:llmsamples3}, and \ref{tab:llmsamples4}). Using GPT-4o-mini as the base LLM model, this is done in three distinct stages. First, given information about the user, a traditional machine learning model predicts individuals’ preferred news sources and rhetorical style. Next, an external database, containing over 70,000 full-length articles, is queried for articles pertaining the statement. We specifically focused on 10 news sources, balancing representation from left, right, and center ideological leanings. The model then performs a summarization step, prompted to pull in relevant information from the resulting query. Finally, the model is instructed to reframe the summarized information to incorporate the predicted rhetorical style.

\subsection*{Results}

\subsubsection*{Personalized LLM Responses Shift Beliefs Toward Factual Accuracy}

\begin{figure}[h] % Do not use \begin{figure*}
	\centering
	\includegraphics[width=0.95\textwidth]{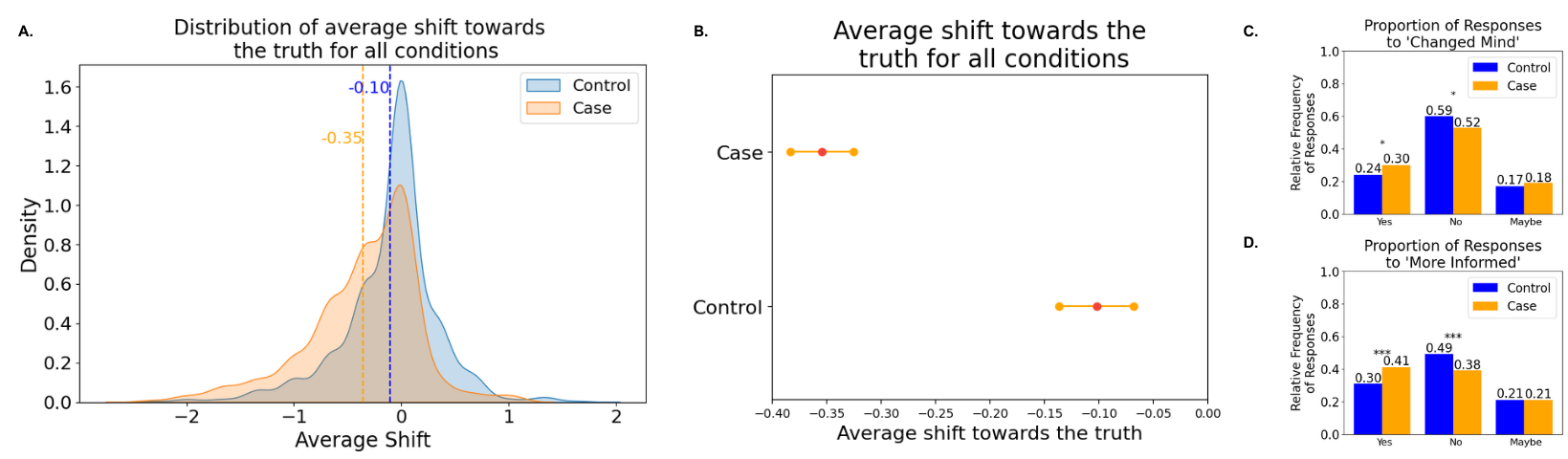} % for an image file named example_figure.*
	% Pick an appriopriate width for the size of the image

	% Captions go below figures
	\caption{\textbf{Individuals in the treatment condition move towards the truth.} \textbf{Panel A} depicts the distribution of the average shift towards accuracy for both treatment and control, where the negative direction represents moving towards accuracy. \textbf{Panel B} depicts the confidence intervals of the two distributions with 95\% confidence. The intervals are non-overlapping, implying significance between the distributions. \textbf{Panel C} depicts the relative frequencies of the user responses to having changed their minds through the course of the experiment. The treatment condition shows more individuals responding ``Yes'' and fewer individuals responding ``No'' and are significant with p $<$ 0.05.\textbf{ Panel D} depicts the relative frequencies of user responses to having felt more informed after going through the experiment. The treatment condition shows more individuals responding ``Yes'' and fewer individuals responding ``No'' and are significant with p $<$ 0.001. These results cumulatively indicate the intervention had a tangible impact on belief update towards the ground truth.}
    \label{fig:beliefupdate} 
\end{figure}

Belief shift was computed by centering each participant’s response score around the veracity label (e.g., true or false) and measuring the signed change in their updated rating, where a negative shift implies a shift towards the accuracy (see SM Section \ref{evaluation_metrics}). We find that participants who received personalized responses from the LLM bot were significantly more likely to update their beliefs in the direction of factual accuracy, compared to those in the control condition. As shown in Figure~\ref{fig:beliefupdate} (Panels A and B) and Table~\ref{tab:overallshift}, the average belief shift was significantly greater in the treatment group as a result of the bot's presence, with an average shift of -0.35 in the treatment and -0.1 in the control ($p < 0.0001$; see SM Section \ref{robustness_supp} and Figure~\ref{fig:stage2_beliefupdatereason}). This pattern held across statement factuality, salience, economic relevance, and self-reported media skepticism (see SM Section \ref{studylevel}, Figures~\ref{fig:finding3_split_truevfalse} - \ref{fig:finding3_split_newstrust}, and Tables~\ref{tab:robustness} and \ref{tab:ttesttrustsome}). It also held across data collection periods (pre-election vs. post-election; see SM Section \ref{prepost}, Figures~\ref{fig:alldist_supplementals} - \ref{fig:case_poltrust}, and Tables~\ref{tab:controlpvp} - \ref{tab:caseproofofinsig}), and demographic groups (see SM Sections \ref{belief_demographic} and \ref{belief_demographic_regress}, Figures~\ref{fig:misinfo_diff_age} - \ref{fig:misinfo_diff_pol_strength}, and Tables~ \ref{tab:between_group_analysis_education} - \ref{tab:between_group_analysis_pol_strength}, Tables~\ref{tab:misinfo_diff_control_reg} and \ref{tab:misinfo_diff_case_reg} for regressions). This is consistent with hypothesis 1.

Self-reported outcomes from the post-survey after completing the experiment further supported the intervention's effect. Participants in the treatment group were more likely to say they changed their mind about any of the statements (Figure~\ref{fig:beliefupdate}, Panel C; $p < 0.05$) and more likely to report feeling better informed after the experiment compared to those in the control (Panel D; $p < 0.001$; see SM Table~\ref{tab:proportiontestsmain} for proportion test results). Our findings were consistent across both time-frames (see SM Section \ref{prepost}, Figures~\ref{fig:cont_proportions} and \ref{fig:case_proportions}, and Tables~\ref{tab:controlproofpropinsig} and \ref{tab:caseproofpropinsig}). These results suggest that personalized LLM messaging influenced belief revision as well as  participants’ confidence about being informed.

%%%%%%%%%%%%%%%%%START HERE%%%%%%%%%%%%%%
\subsubsection*{Participants Follow the Bot and Seek Out Others Aligned with Its Beliefs}

\begin{figure}[htbp] % Do not use \begin{figure*}
    \centering
    \includegraphics[width=0.95\linewidth]{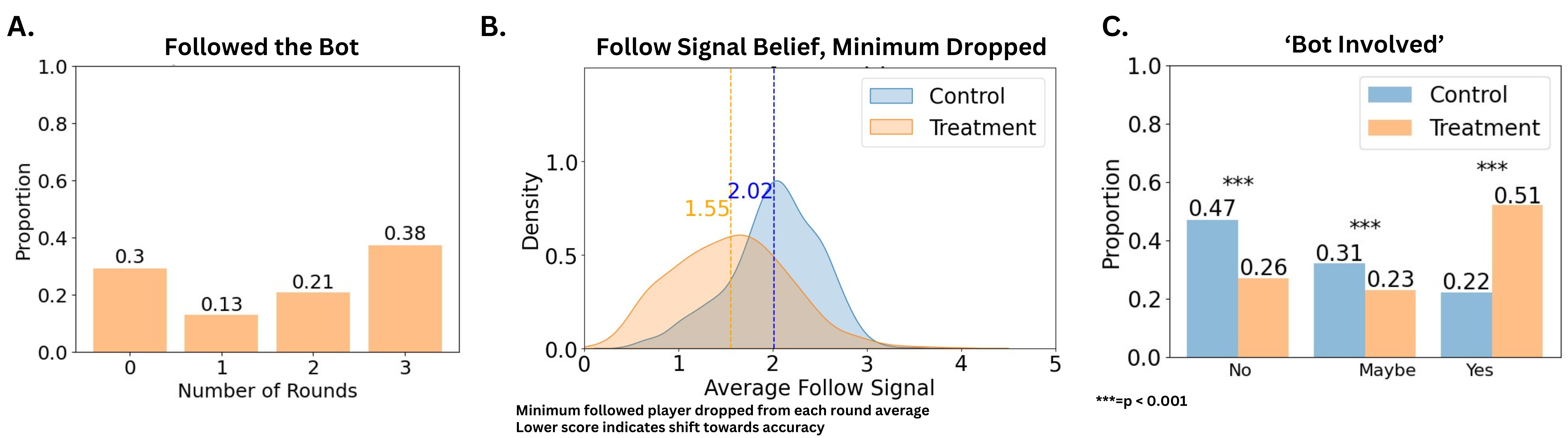} 
	% Captions go below figures
	\caption{\textbf{Individuals in the treatment condition choose to follow the bot and others with similar opinions.} \textbf{Panel A} depicts a histogram of the number of rounds players followed the bot. Most individuals followed the bot for at least one round, and when individuals followed the bot, they tended to continue to follow the bot over the course of the whole experiment (three rounds). \textbf{Panel B} shows the average follow signal of the treatment and control with the minimum removed from each round average, where 0 represents the ground truth. When the bot is present, the bot will be removed as it will always suggest the ground truth. The mean of the treatment distribution is significantly closer to the truth than the control condition. This shows that, even with the bot removed, those in the treatment chose to follow others whose beliefs were more accurate, compared to those followed in the control.\textbf{ Panel C} depicts the relative frequency of user responses to whether a bot was involved in the study. More participants in the treatment selected ``Yes'' and fewer participants in the treatment selected ``No'' and ``Maybe''. All three findings are significant at p \textless 0.001. These figures cumulatively show that individuals within the treatment chose to follow the bot and others with beliefs similar to the bot, despite awareness of the intervention.}
    \label{fig:networkupdate}
\end{figure}

The LLM intervention influenced the structure of participants’ social networks. Consistent with hypothesis 2a, the majority of participants in the treatment condition (70.2\%) chose to follow the bot at least once, and over half of these individuals (54.4\%) continued following it for all three rounds of the study (Figure~\ref{fig:networkupdate}, Panel A; see SM Sections \ref{regress_botfol} and \ref{prop_botfol}, Figures ~\ref{fig:follow_age} - \ref{fig:follow_pol_strength} and Table~\ref{tab:multilinear_reg} suggesting no significant difference in results between demographic groups). While the effect of the bot is prevalent throughout the treatment condition results, the use of personal pronouns by participants in their responses and a mix of source-leaning cues by the bot responses lead to stronger shifts (see SM Section \ref{persuasivellm} and Figures~\ref{fig:botpersuasion_personal} and \ref{fig:botpersuasion_sourcelean}). We also find that participants who follow the bot in stage 3 tend to shift their beliefs closer to the truth in stage 2 (see SM Section \ref{botfolvchange}). This suggests that participants did not treat the bot as a one-off novelty, but rather as a sustained source of potential guidance within the evolving network.

Beyond the bot itself, participants tended to follow human peers whose views were more closely aligned with the bot’s accurate stance, as predicted by hypothesis 2b. While both groups engage in truth-seeking behaviors \cite{pennycook2021shifting, bond2023engagement} (see SM Section \ref{networklevel} and Figure \ref{fig:botfollowvrandomtruth}), we find individuals in the treatment experience this truth pursuit more dramatically than those in the control. To quantify this tendency, we computed a ``follow signal'' for each participant: the mean factual accuracy score (calibrated by truth direction) of the individuals they chose to follow at each round, averaged over all three rounds. The score was normalized such that responses closer to the truth yielded lower values, and the most truthful individual---either the bot, if followed, or the lowest-scoring peer---was excluded to enable fair comparisons (see SM Section \ref{evaluation_metrics} and Equation~\ref{eq:followsig}). The average follow signal was significantly lower in the treatment condition (1.55) than in the control (2.02; $p < 0.0001$), as shown in Figure~\ref{fig:networkupdate}, Panel B and Table~\ref{tab:followsignal}. Given the adjusted belief scale ranges from 0 (most accurate option) to 4 (least accurate option), our results suggest that participants in the treatment notably shift towards more truthful network compositions. Moreover, we find that networks became more assortative between stages 2 and 3 (see SM Section \ref{networklevel} and Figures~\ref{fig:assortativity_control} and \ref{fig:assortativity_case}), which implies individuals are not simply choosing who they follow at random. Thus, the presence of the bot had downstream consequences in shaping individuals’ subsequent social networks --- here toward more accurate information.

Importantly, participants were not unaware of the bot's presence. In the post-surveys, 51\% of treatment participants reported interacting with a bot, compared to 22\% in the control group (Figure~\ref{fig:networkupdate}, Panel C). Despite this recognition, the bot remained influential \cite{matz2024potential}. Participants in the treatment adopted similar beliefs and social connections, including following the bot, suggesting that perceived artificiality did not undermine trust or effectiveness \cite{matz2024potential}. This persistence of influence, even under partial awareness, reinforces the robustness of the intervention's social effects.

\subsection*{Discussion}
We show that personalized LLM interventions can support belief updates toward ground truth in the context of political statements. The fact that participants followed the bot at high rates and sought out others aligned with its position indicates that targeted AI agents can influence both individual cognition and network-level structure. Individuals tend to seek truth \cite{pennycook2021shifting}, and by embedding a truthful agent within the social context, our intervention amplified this baseline tendency. We find no difference in the efficacy of the intervention across multiple factors, including the statement veracity, economic relevance, topic salience, individual demographic traits, and time of data collection. This suggests that the intervention is robust. 

To our knowledge, this is the first study to empirically demonstrate that a personalized LLM embedded in a social network can reshape the surrounding social topology in a direction aligned with factual accuracy. Participants in the treatment condition were more likely to connect with others whose views were closer to the ground truth, compared to those in the control group. These findings point to the dual function of such agents: they serve as direct informants and as catalysts for reconfiguring information flows in a network. In contemporary online ecosystems, where individuals are routinely exposed to conflicting signals, personalized LLMs may offer a scalable way to redirect attention toward reliable sources and re-anchor belief trajectories.

Notably, the observed belief shifts occurred even without a back-and-forth dialogue and even when participants were not explicitly aware that they were interacting with a bot (note that they were made aware that a bot might be present and had provided consent to interacting with one). These features make our experimental setup more reflective of actual digital environments, where users often engage with brief messages and may only have partial awareness of whether content is AI-generated.

At the same time, the mechanisms that enable this influence also carry clear risks. Prior studies have shown that users often accept LLM responses, even when they are factually incorrect or deliberately misleading \cite{sharma2024would, deverna2024fact}. Our results underscore how even modest personalization and contextual awareness can enhance credibility. Thus, replacing our guardrails with maligned structures makes it plausible that a similarly structured but malicious bot could amplify misinformation rather than mitigate it. As LLM-driven systems become more integrated into public discourse, designing effective guardrails such as transparent sourcing, regular audits, and limits on persuasive framing will be essential. Our findings help define what such systems can do and where oversight must be targeted. The results make clear that LLMs have enormous potential in not only altering beliefs but also altering the very social fabric in which people interact. 

As moderating content on large-scale social media platforms becomes more difficult to complete directly, personalized LLMs could be utilized as a potential method. With user permission, individuals could receive persuasive, factual responses to potential sources of misinformed beliefs in any information domain. To ensure responsible use, not only should guardrails as we describe here be considered, but also be extended to further protect users and incorporate auditing, such as receiving consent and not collecting any personally identifiable information. Such personalized content should focus on ensuring transparency, fairness, and accountability in how the system operates -- positioning them as assistive technologies for navigating today’s complex information environment.

This work can shape future research within the space of personalized LLMs and factors of belief. For example, conducting a similar study within other information environments would help reinforce the generalization of these findings. Furthermore, investigating the effects of such an intervention in real-world platforms would shed light on its efficacy in a setting with more complex information streams. While considerable research addresses how LLMs affect beliefs, our results accentuate that the impact may go further by altering the social networks within which we exchange information. Thus, it is imperative to continue to study LLM interactions within social settings consisting of both human and AI interactions, and to protect the end users throughout the process.

%%%%%%%%%%%%%%% MAIN TEXT FIGURES %%%%%%%%%%%%%%%

\clearpage

%%%%%%%%%%%%%%%% MAIN TEXT TABLES %%%%%%%%%%%%%%%

\begin{table}[h]
    \centering
    \caption{\textbf{Mean, Standard Deviation, and Confidence Intervals of Treatment and Control Average Belief Shifts}. Confidence intervals are non-overlapping at $\alpha$ = 0.05, which implies significance.}
    \label{tab:overallshift}
    \begin{tabular}{|c|c|c|c|}
        \hline
         &  Mean & Standard Deviation & Confidence Interval\\ \hline
        Treatment & -0.102 & 0.454 & (-0.137, -0.067)\\ \hline
        Control & -0.354 & 0.517 & (-0.394, -0.315) \\
        \hline
    \end{tabular}
\end{table}

\begin{table}[h]
    \centering
        \caption{\textbf{Mean, Standard Deviation, and Confidence Intervals of Average Follow Signals.} Confidence intervals are non-overlapping at $\alpha$ = 0.05, which implies significance.}
    \label{tab:followsignal}
    \begin{tabular}{|c|c|c|c|}
        \hline
         & Mean & Standard Deviation & Confidence Interval \\ \hline

        Treatment Follow & 1.554 & 0.613 & (1.507, 1.602) \\ \hline

        Control Follow & 2.016 & 0.482 & (1.979, 2.053) \\
        \hline
    \end{tabular}
\end{table}

%%%%%%%%%%%%%%%% REFERENCES %%%%%%%%%%%%%%%

\clearpage % Clear all remaining figures and tables then start a new page

\bibliography{main} 
\bibliographystyle{sciencemag}

%%%%%%%%%%%%%%%% START OF SUPPLEMENT %%%%%%%%%%%%%%%

% Figures, tables, equations and pages in the supplement are numbered S1, S2 etc.
\renewcommand{\thefigure}{S\arabic{figure}}
\renewcommand{\thetable}{S\arabic{table}}
\renewcommand{\theequation}{S\arabic{equation}}
\renewcommand{\thepage}{S\arabic{page}}
\setcounter{figure}{0}
\setcounter{table}{0}
\setcounter{equation}{0}
\setcounter{page}{1} % not 0 as \newpage already started a supplementary page
% References continue the numbering from the main text.

%%%%%%%%%%%%%%%% SUPPLEMENT TITLE PAGE %%%%%%%%%%%%%%%
\newpage 
\begin{center}
\section*{Supplementary Materials for\\ \scititle}

Adiba Proma$^{\ast \dagger}$, 
Neeley Pate$^{\dagger}$, 
Sean Kelty, \\
Gourab Ghoshal, 
James N. Druckman,  
Ehsan Hoque \\
% Additional lines of authors should be inserted using the \and command (not \\)
% Institution list, in a slightly smaller font

% Identify at least one corresponding author, with contact email address
\small$^\ast$ Corresponding author. Email: aproma@cs.rochester.edu \\
% Joint contributions can be indicated like this
\small$^\dagger$These authors contributed equally to this work.

\end{center}

% Fill out the numbers for each type of supplementary material,
% and delete any lines that aren't applicable.
% These are just example numbers that don't match the rest of this template.
\subsubsection*{This PDF file includes:}
Sections S1 to S4 \\
Figures S1 to S35\\
Tables S1 to S35\\

% \subsubsection*{Other Supplementary Materials for this manuscript:}
% % LLM S1\\
% Data S1 to S2

\newpage

%%%%%%%%%%%%%%%% MATERIALS AND METHODS %%%%%%%%%%%%%%%

%%%%%%%%%%%%%%%% SUPPLEMENTARY TEXT %%%%%%%%%%%%%%%
%\subsection*{Supplementary Text}
% The Supplementary Text section can only be used to directly support statements made in the main text
% e.g. to present more detailed justifications of assumptions, investigate alternative scenarios,
% provide extended acknowledgements etc.
% Material in this section cannot claim results or conclusions that weren't mentioned in the main text.
% To refer to this section from the main text, just write (Supplementary Text).

%%BEGIN NEELEY RESTRUCTURING%%
\section*{Materials and Methods}
\label{methods}
We are interested in how personalized information from an LLM bot (relative to the absence of a personalized bot) impacts beliefs and subsequent social network construction. The basic structure of the study placed individuals in a social network and then had them: 1) report their belief about the veracity of a statement and explain their rationale with brief text, 2) learn about the veracity judgments of others in the social network, 3) have the opportunity to update their belief and explain it again, and, 4) choose whom from the network they would like to follow. We made two additional design decisions. First, we followed prior political science work on opinion formation by including four entities in a social network \cite{klar2014partisanship, druckman2018no}. Participants in each network took part concurrently (and were recruited to do so). There was a median of 16 participants per study batch, within the range 10-28 participants. This variation is due in some part to the timing in the day of the experiment and participant availability. Framework limitations also play a role, such as 10 participants being required to participate in order for the experiment to work properly, and a maximum of 24 people can be supported in the treatment because of time constraints on bot response generation. Second, we created a dynamic setting where individuals took part in three rounds (with each proceeding as described below). A choice in one round influenced later rounds. Specifically, when a participant chose someone to follow in an earlier round, that actor became more likely to appear in the participant’s subsequent round(s) of the social networks. This increased experimental realism, which refers to the extent to which participants take the study seriously \cite{druckman2022experimental}. Making consequential choices presumably increases engagement.

For reference, see Figures \ref{fig:stage1}, \ref{fig:stage2}, \ref{fig:stage3} to view the interface for all parts of the experiment, which are described below. Note that while these figures depict real responses from participants and a bot, this does not reflect a real snapshot of one round of a given experiment; the figures are purely for framework description. While the interfaces remain consistent across all rounds, the behavior in round 1 differs slightly from rounds 2 and 3.

The precise sequence for the \textbf{first round} was as follows.

\subsection*{\textbf{Stage 1: Response stage (First round)}}
Participants are provided with a prompt statement on one of the following topics: immigration, inflation, fuel and oil, or overpopulation. Some of these statements were true while others were false. More details on how these statements were selected are provided in section \ref{PoliticalStatement} of Supplemental Materials. Participants reported how much they believed the statement to be true on a 5-point scale (with higher scores indicating stronger belief that the statement is true). They additionally provided reasoning for their rating in a maximum of 1,000 characters. We refer to each evaluation and its accompanying explanation as a ``response entry.'' All participants in a given network receive the same statement. 

\subsection*{\textbf{Stage 2: Revision stage (First round)}}
In the revision stage, participants received the response entries of a few others in the network. Participants re-rated the same prompt again on a 5-point scale, and also selected from a pre-determined list of reasons the rationale behind their selection (see Supplement Section \ref{selfreportbeliefupdate} for options). The participant's initial rating is displayed to ensure that any change in their rating is intentional. However, even if they do not change their response, participants still had to click the scale manually. 

This stage includes our experimental manipulation. In the control, participants were shown the entries of three other participants in the study. In the treatment, participants viewed the entries of two other participants as well as that of the personalized bot. The personalized LLM bot always provided the correct value, with a response entry tailored to the participant’s preference. 

\subsection*{\textbf{Stage 3: Selection stage (First round)}}
In the selection stage, participants were presented with six response entries. Participants learned these entries came from those they previously followed, and/or those `recommended' by the system. Participants saw the three peers from stage 2 (including the bot for the treatment condition), with the other three being randomly recommended. This setup ensures that participants get access to diverse responses. The only difference in the selection stage between control and the treatment is that in the treatment, one of the response entries is that of the personalized bot. 

Participants must then like or dislike each response entry (to prompt thoughtful engagement with the content) and select (i.e., ``followed") three peers whose answers they would like to subsequently see in the next round. They could stick to the peers assigned at stage 2 or pick new ones (or do a mix of both). Through this stage, we can evaluate how participants restructured their networks, including whether they chose to follow the personalized bot and incorporate the bot into their networks. We can also quantify the changes in the participants' networks given the presence of the bot. 

\vspace{2em}
\noindent

\textbf{Rounds 2 and 3} were similar, with a few changes to account for increasing experimental realism. Participants have the same stage 1 but with a new prompt on the same topic (i.e., a new statement on the same issue). Stages 2 and 3 for rounds 2 and 3) differed, however, as follows. 

\subsection*{\textbf{Stage 2: Revision stage}}
Similar to round 1, participants saw three others. For the control, there was one randomly chosen person that they had followed in the prior round and two others they had not followed (randomly chosen). For the treatment, the participants saw the personalized bot, a followed human participant (chosen at random), and a non-followed human participant (chosen at random). As mentioned, we ensured participants saw at least one person they had followed to provide them with partial autonomy over their networks (and thus, their social influence). We included non-followed individuals to minimize the possibility of extreme homophily. 

\subsection*{\textbf{Stage 3: Selection stage}}
In stage 3, participants saw the peers they were shown in stage 2 (including the bot for the treatment condition), those they followed in the previous round, and the rest were randomly recommended. This ensures that the setup maintains the ``social media" structure, where once someone follows a person, they do not get unfollowed unless the user manually unfollows them. 

\subsection*{\textbf{Pre and Post Survey}}

After users provide their consent to be a part of the study, we collect a few demographic and personality measures. For gathering demographic data, we used the categories as used in previous literature \cite{proma2024exploring}. For measuring personality, we utilized shortened ``big 5'' personality trait questionnaire which contained 10 questions \cite{rammstedt2007measuring}. This provides the basis for our personalized LLM and allows us to check demographic differences in behavior. Upon entering the study, the users provide the following information:

\begin{itemize}
    \item Username
    \item ProlificID
    \item Age (18-24, 25-34, 35-44, 45-54, 65+)
    \item Gender (Male, Female, Non-binary, Prefer not to answer)
    \item Race (White, Hispanic, Black or African American, Asian, American Indian or Alaska Native, Other)
    \item Education level (High school diploma, Associate degree, Bachelor's degree, Graduate or professional degree, Other)
    \item Political Party (Democratic, Republican, Independent, Other)
    \item Strength of Political Identification (Very weak, Somewhat weak, Neutral, Somewhat strong, Very strong)
    \item Location (State)
    \item Occupation
    \item Extroversion
    \item Agreeableness
    \item Conscientiousness
    \item Neuroticism
    \item Openness
\end{itemize}

Upon completing all three rounds of the study, users then fill out a post survey survey, which includes the following questions (those bolded considered for analysis within the manuscript):

\begin{itemize}
    \item How often do you get your news from social media?
    \item Please identify which social media platforms you get your news from: (check all that apply)
    \item Please identity which news sources you generally trust: (check all that apply)
    \item Which of the following information characteristics do you consider when determining its factuality? (check all that apply)
    \item Which of the following personal characteristics have an impact when determining information factuality? (check all that apply)
    \item \textbf{Did going through the experiment change your mind about any statements?}
    \item \textbf{Do you feel more informed about the political topics shown after going through the experiment?}
    \item \textbf{How do you feel about political news at this current point in time?}
    \item When considering the responses of others, how relevant were any sources they cited in your decision making?
    \item When considering the responses of others, how relevant was the logic of their statement in your decision making?
    \item When considering the responses of others, how relevant was the emotion of their statement in your decision making?
    \item \textbf{At any point of the experiment, did you feel like you were interacting with bots instead of real participants?}
    \item Please provide your feedback on the experience. This can include any problems you encountered or any other issues you faced during the experiment.
\end{itemize}

Upon submitting the post survey, participants were then debriefed regarding statement factuality and the potential for bot engagement during the study (note that participants were previously told that there may or may not be a bot present). They were also provided with contact information of the researchers in the treatment they had further questions.

\section{Evaluation Metrics}
\label{evaluation_metrics}

To generate all shift plots (i.e., plots showing shift towards or away from the truth),  we first encoded the initial and updated scores as a value from [0, 4], where 0 represents ``Strongly Disbelieve'' and 4 represents ``Strongly Believe,'' since these were the labels provided to participants during the study. To account for the distinction between true and false statements,  the difference between the two values for statements with the ground truth of ``true'' are calculated as initial - updated, and statements with the ground truth ``false'' are calculated as updated - initial. These calculations ensure that a shift in the intended direction is encoded as a negative change, while a shift towards misinformed beliefs is encoded as a positive change. For each person, this difference was averaged over all rounds, and then plotted using a kernel density plot. For determining distribution differences, all calculations use the t-score to calculate the confidence interval and p-value. When confidence intervals between findings of interest overlap, Levene’s test is utilized to determine whether the variance between the two samples are equivalent, then the appropriate t-test is conducted based on those results to determine significance. Levene’s test is considered to be a robust test \cite{Boos_Bartlett_Boos_Box_Brown_Brown_Conover_Draper_Geng_1999} utilized in prior works \cite{Sun_Zhou_Hauptschein_Rappoport_Xu_Navarro}. We use t-tests because the standard deviation of the underlying population is not known, and initial tests indicate the distribution was not normal.

For calculating the follow signal, we consider the subset of individuals within a given round that were followed by each participant. For each person's followed subset, we drop the most truthful person when 3 people were followed (which is mostly the case, except for a few edge cases). We then average the scores of the followed participants, again normalized to true/false statements by considering the initial score in the false case and 4 - initial score in the true case for all followed players. This ensures a score of 0 is the most truthful score. After computing the average follow signal for each round, we then find the mean across all rounds; this represents the average follow signal across the whole experiment for a given individual (see Equation \ref{eq:followsig}). This is again plotted using a kernel density plot and tested for significance in a similar procedure as described for shift evaluations.

\begin{equation}
    \delta_{follow signal} = \frac{\sum^3_{i = 1} \frac{(\sum^n_{j = 1} f_j) - f_{min}}{n-1}}{3}
    \label{eq:followsig}
\end{equation}

When computing proportions, we take the number of participants who fall into a specified category and divide it by the sum of all answers given. To compare proportions, we utilize a two proportion z test, where we compare the proportion of the given category across two conditions.

\section{Designing the Experiment and the Personalized LLM Bot}
\label{exp_llm_design_supp}
Below describes our specific processes for determining elements of experiment design and the personalized LLM bot.

\subsection{Experimental Design}
\label{exp_design_supp}

Beyond the experimental platform design, other aspects of the design had to be considered: the news sources the LLM bot should be able to access, the political statements individuals would see during the experiment, and the selection of our time-frames for hosting these studies. Each of these is discussed in detail below.

\subsubsection{Selection of the news sources}
\label{news_sources}
The news sources an individual chooses to trust can be affected by a variety of factors, and as a result, our goal was to select a wide variety of news sources that would encapsulate the potentially diverse opinions/preferences of individuals. Several resources were consulted and compared to create a comprehensive, representative list of trustworthy sources. One of the most important factors for trust in news sources is political affiliation. According to a 2020 report from Pew Research Center \cite{jurkowitz2020democrats}, a large proportion of U.S. adults trust sources such as CNN, CBS, and Fox News. Similarly, a Cooperative Election Study (CES) Common Content 2022 study conducted ``across a nationally representative sample of 60,000 American adults" which showed that Americans engage with several news networks, including CNN, MSNBC, and Fox News \cite{schaffner2021cooperative}. Next, we considered AllSides Media Bias Chart and ``The Chart" created by Ad Fontes in 2018 \cite{harvardResearchGuides} to determine the political leaning and the factuality of the sources. The factuality and leaning of these sources were cross-referenced with those provided by ground.news \cite{GroundNews}, a website that also provides media bias and factuality ratings aggregated according to the news reported by a news source. Through this process, we shortlist potential news sources, ensuring a relatively equal representation of left, right, and central sources. Next, we checked whether their API is publicly available through GNews and if enough articles can be extracted from that source to build our RAG. Our final set of 10 news sources were Washington Examiner (right), New York Post (right), Fox News (right), AP News (center), The Economist (center), BBC (center), New York Times (left), CNN (left), Washington Post (left), and MSNBC (left). Note that the news preference dataset was built considering these 10 news sources, and the articles from these news sources were used to build our RAG. The number of articles extracted from each source is provided in Table \ref{tablearticles}.

\subsubsection{Political Statement Generation}
\label{PoliticalStatement}
We selected four specific topics --- overpopulation, inflation, immigration, and fuel and oil --- of varying saliency levels, to ensure generalizability of our experimental setup. The topics were selected based on reported saliency of issues during the period of the 2024 US presidential election, balancing with two high salience (inflation and immigration) and two low salience (overpopulation and fuel and oil) topics \cite{gallup, Jones_2025}. We curate a series of political statements for each of the topics. These statements must be bipartisan and with objective ground truth. The statements themselves can be true or false. To ensure a variety of statements, we generate them in two different ways. First, we conduct a series of in-person focus group studies, where participants come up with their own politics-related statements, and find their corresponding verifying information online. This provided insights into the personal aspects of misinformation (i.e., what makes statements believable for individuals). To create the statements used for this study, we relied on credible information to prove or disprove our statements from government published statistics, Politifact, FactCheck.org and Gallup. The final selected statements are provided in Tables \ref{statements_part1}, \ref{statements_part2}, and \ref{statements_part3}. 

% All the participants are shown the statements and the corresponding verifying information generated, through a majority voting system, the top ones are selected to be included. Second, we look through online sources such as Politifact, AP News (specifically, a column called Not Real News) and FactCheck.org to find the most popular topics and corresponding verifying information. 

\subsubsection{Selection of our time-frames}
\label{timeframe}
We conducted our study before and after elections. The duration for before elections was October 1 - October 31, 2024 and after the elections, it was November 15 - December 15, 2024. There are prior work suggesting that there is some change in beliefs pre and post elections \cite{banducci2003elections}. However, we did not have prior predictions about it moderating the treatment effects. Our results show that the differences in context do not, in fact, affect the effectiveness of the intervention. More details about these results are presented in Section \ref{prepost}.

% However, to our knowledge, there are no prior predictions about it moderating the treatment effects. Rather, our selected time frames were chosen to see if the results are robust and generalizable across contexts. We argue that pre-election and post-election contexts are different since there is a clear ``winner" and a clear ``loser" of the elections, which can affect individual perceptions \cite{fasching2024persistent, banducci2003elections}. Moreover, various studies have also shown that elections act as a short-term polarizing event, with partisan identities becoming less salient after the elections, further contributing to the fact that the context between pre and post elections are different \cite{michelitch2018electoral}. 

% NOTE -- UMAP PAPER SHOULD GO HERE WHEN IT COMES OUT FINALLY 

\subsection{Personalized LLM Bot}
\label{personalizedllmbot}
Our goal was to first design an LLM bot that can incorporate an individual's preferred news sources and rhetorical styling to create compelling arguments. The personalized bot design follows a four-step pipeline to generate the responses using ChatGPT as the base model. At first, we predict users' news source preferences and their preferred rhetorical style from ethos, pathos, and logos using K-Means clustering and collaborative filtering. The preferred news source predictions are then passed to the RAG model. The RAG contains news articles from 2023 to 2024 from the 10 sources previously mentioned in section \ref{news_sources}. The model finds related articles, considering the predicted news source preference, and then summarizes those articles to 3-4 sentences. The exact prompts used are provided below. This summarized information, along with the predicted rhetorical style, is then passed to another LLM, which rephrases the summarized information to incorporate the rhetorical style provided. We provide tables that capture a variety of LLM bot final outputs, along with their initial input statement (Tables \ref{tab:llmsamples1},\ref{tab:llmsamples2}, \ref{tab:llmsamples3},\ref{tab:llmsamples4}). For anonymity, the demographics and personality input into the system are not given. The table shows the different variations in facts, sources, and styles the LLM utilized.

% \subsubsection*{Prompts used in the LLM}
% \label{prompts}
% Listed below are all of the prompts used to generate the LLM responses. The LLM is asked to generate two items. First, the LLM generates a summary of the full-length news articles provided to it. Then, the LLM is asked to rephrase the summary using rhetorical styling. Here, ``Information'' refers to the statement to prove or disprove, ``Excerpts'' refers to the full-length news articles to support or disprove the ``Information'', and the ``Summary'' refers to the output of the summarization generation.

% \subsubsection{Questionnaire Utilized in Creation of Preference Dataset}

% \textcolor{red}{In order to identify personal preferences for news source and rhetorical styling, we previously collected data from individuals ...}

\subsubsection{Summarization Generation Prompts}

\textbf{True Statement Condition}: Given that the `Information' is true, summarize the relevant facts of each of the `Excerpts' that support the `Information'. Keep the source for each of the `Excerpts' in the summary. After summarizing, remove all summaries not relevant to `Information'. Do not use any hate speech or vulgar language.

\noindent
\textbf{False Statement Condition}: Given that the `Information' is false, summarize the relevant facts of each of the `Excerpts' that disprove the `Information'. Keep the source for each of the `Excerpts' in the summary. After summarizing, remove all summaries not relevant to `Information'. Do not use any hate speech or vulgar language.

\subsubsection{Rhetorical Styling Prompts}
\textbf{True Statement Condition}: You are an informed citizen, persuading another citizen that the `Information' is true. Read the `Summary' and identify relevant facts. Then, write a first person response to prove the `Information' by \textit{$<$best tactic here$>$}. Do not use hate speech or vulgar language. Limit the response to 4 sentences.

\noindent
\textbf{False Statement Condition}: You are an informed citizen, persuading another citizen that the `Information' is false. Read the `Summary' and identify relevant facts. Then, write a first person response to disprove the `Information' by \textit{$<$best tactic here$>$}. Do not use hate speech or vulgar language. Limit the response to 4 sentences. 

\textbf{Ethos Tactic}: emphasizing the credibility of the news sources.

\textbf{Pathos Tactic}: emphasizing empathy towards the reader

\textbf{Logos Tactic}: emphasizing the impact of the evidence.

% \subsection{Personalized LLM Bot Final Output Samples}

% Provided here is a table of a variety of LLM bot final outputs, along with their initial input statement (Tables \ref{tab:llmsamples1},\ref{tab:llmsamples2}, \ref{tab:llmsamples3},\ref{tab:llmsamples4}). For anonymity, the demographics and personality input into the system are not given. The table shows the different variations in facts, sources, and styles the LLM utilized.

\section{Robustness Measures}
\label{robustness_supp}

\subsection{Participant Demographics}
\label{part_demographics}
Table \ref{tab:participant_dem} shows the demographic breakdown of participants in the treatment and control groups. The distributions for control and the treatment are matched in terms of age range, gender, race, political affiliation, and all five personality traits (see Table \ref{tab:chi2_demographics}, \ref{tab:mwu_personality}). 

\subsection{Belief Shift Across Demographic Differences}
\label{belief_demographic}
\begin{figure}
    \centering
    \includegraphics[width=0.7\linewidth]{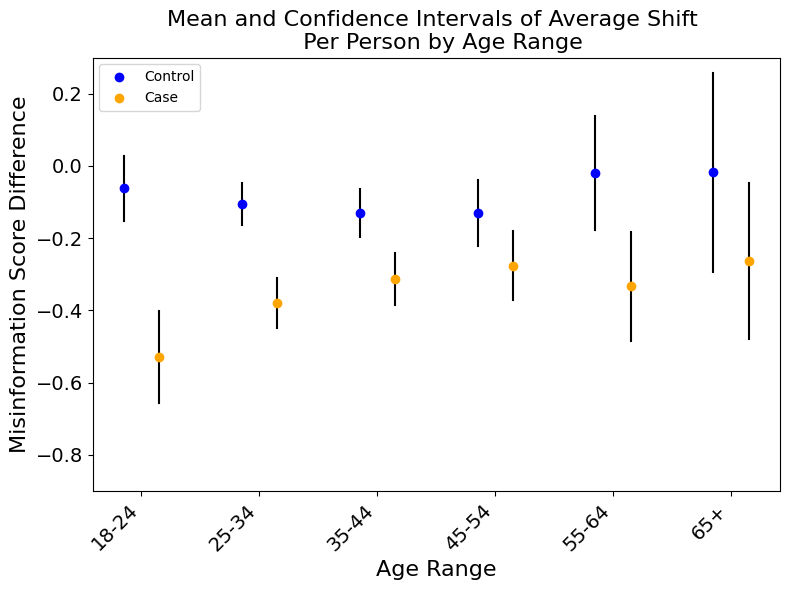}
    \caption{\textbf{Comparison of Mean Misinformation Score Difference for Each Person for both Control and Treatment, Separated by Age Range.} Within-group analysis is not statistically significant, but between-group analysis is statistically significant.}
    \label{fig:misinfo_diff_age}
\end{figure}

\begin{figure}
    \centering
    \includegraphics[width=0.7\linewidth]{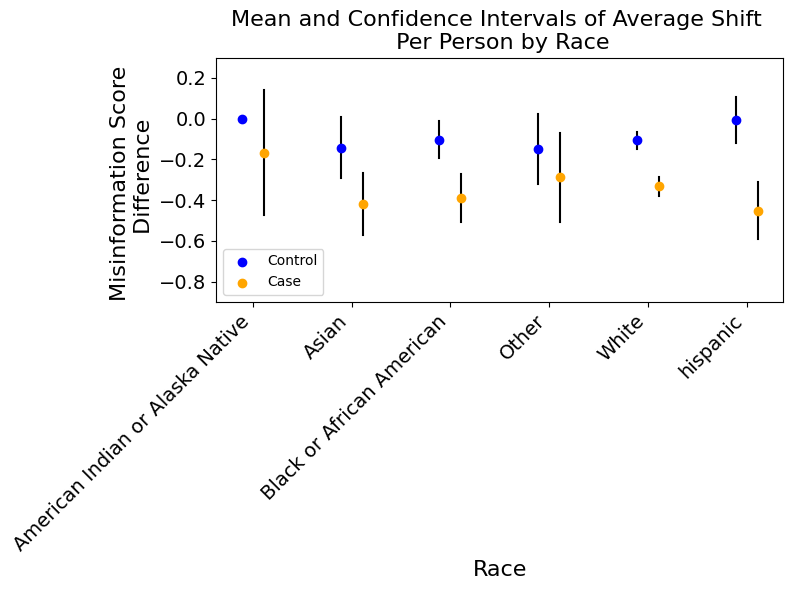}
    \caption{\textbf{Comparison of Mean Misinformation Score Difference for Each Person for both Control and Treatment, Separated by Race.} Within-group analysis is not statistically significant, but between-group analysis is statistically significant.}
    \label{fig:misinfo_diff_race}
\end{figure}

\begin{figure}
    \centering
    \includegraphics[width=0.7\linewidth]{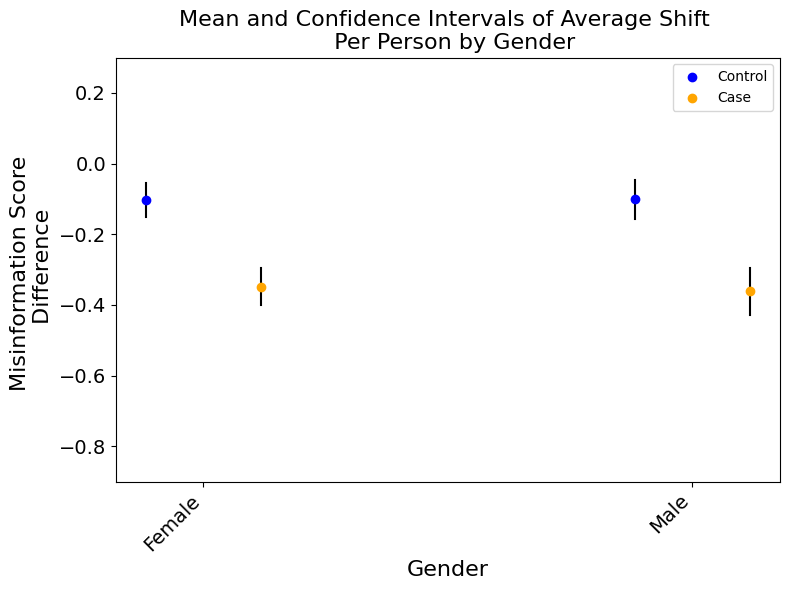}
    \caption{\textbf{Comparison of Mean Misinformation Score Difference for Each Person for both Control and Treatment, Separated by Gender.} Within-group analysis is not statistically significant, but between-group analysis is statistically significant.}
    \label{fig:misinfo_diff_gender}
\end{figure}

\begin{figure}
    \centering
    \includegraphics[width=0.7\linewidth]{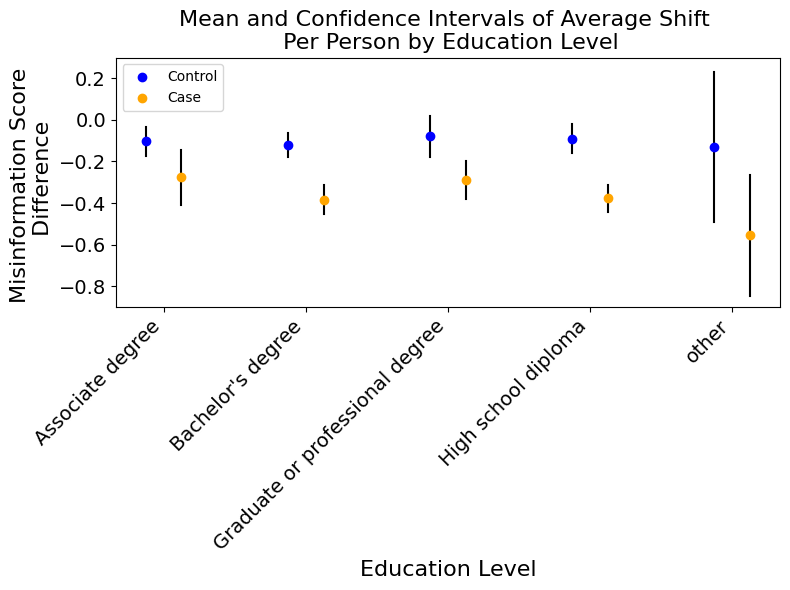}
    \caption{\textbf{Comparison of Mean Misinformation Score Difference for Each Person for both Control and Treatment, Separated by Education.} Within-group analysis is not statistically significant, but between-group analysis is statistically significant.}
    \label{fig:misinfo_diff_education}
\end{figure}

\begin{figure}
    \centering
    \includegraphics[width=0.7\linewidth]{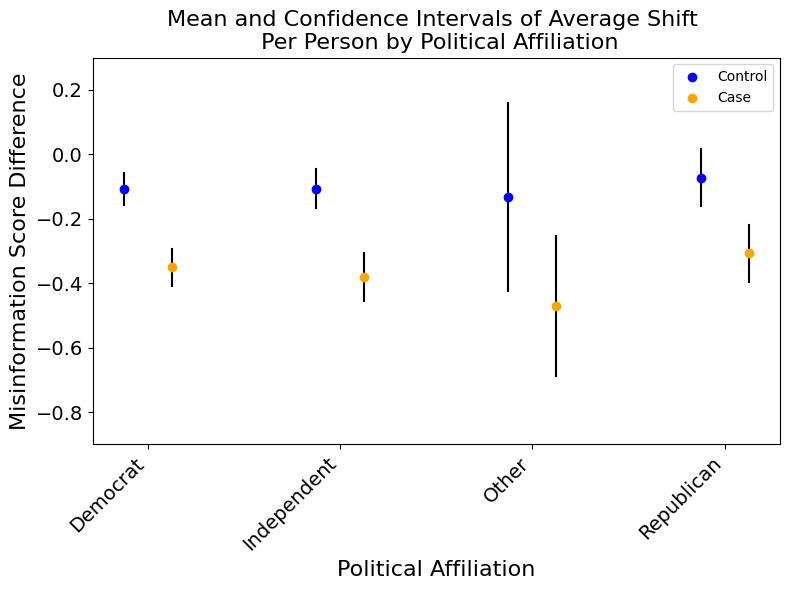}
    \caption{\textbf{Comparison of Mean Misinformation Score Difference for Each Person for both Control and Treatment, Separated by Political Affiliation.} Within-group analysis is not statistically significant, but between-group analysis is statistically significant.}
    \label{fig:misinfo_diff_pol}
\end{figure}

\begin{figure}
    \centering
    \includegraphics[width=0.7\linewidth]{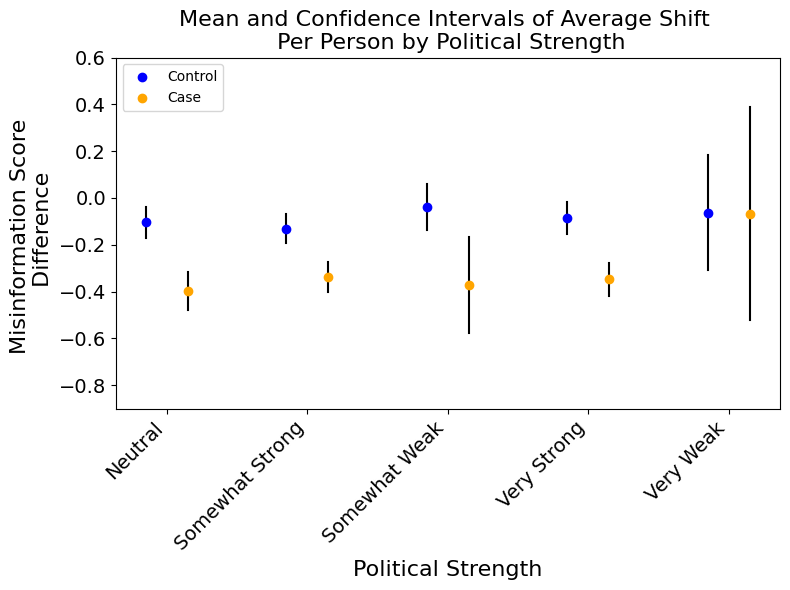}
    \caption{\textbf{Comparison of Mean Misinformation Score Difference for Each Person for both Control and Treatment, Separated by Strength of Political Affiliation.} Within-group analysis is not statistically significant, but between-group analysis is statistically significant.}
    \label{fig:misinfo_diff_pol_strength}
\end{figure}

To further prove robustness of the interventions, we evaluate differences in demographics across participants. We compare average shift (calibrated for true/false statements) for each person, and we find no significant difference within group across any demographics, except for age range in the treatment. Note that older population has been previously shown to be less flexible compared to younger population \cite{wilson2018age}. The plots are shown in Figures \ref{fig:misinfo_diff_age}, \ref{fig:misinfo_diff_race}, \ref{fig:misinfo_diff_education}, \ref{fig:misinfo_diff_gender}, \ref{fig:misinfo_diff_pol}, and \ref{fig:misinfo_diff_pol_strength}. The within analysis using ANOVA are shown in \ref{tab:within_group_analysis_age_range}, \ref{tab:within_group_analysis_race}, \ref{tab:within_group_analysis_education}, \ref{tab:within_group_analysis_gender}, \ref{tab:within_group_analysis_political}, and \ref{tab:within_group_analysis_pol_strength}. The between analysis are shown in Tables \ref{tab:between_group_analysis_age_range}, \ref{tab:between_group_analysis_gender}, \ref{tab:between_group_analysis_political}, \ref{tab:between_group_analysis_pol_strength}, 
\ref{tab:between_group_analysis_race}, and \ref{tab:between_group_analysis_education}. 

\subsection{Regression of Belief Shift Across Demographic and Personality Differences}
\label{belief_demographic_regress}
We use a multiple linear regression model using Strata to model whether there are demography and personality-based differences in if individuals move towards the truth or not in the treatment and the control. All features were not significant in the treatment, except for age range. None of the features in control were significant. Overall, our results suggest that the effect was similar across all individuals. The results are shown in Tables \ref{tab:misinfo_diff_control_reg} and \ref{tab:misinfo_diff_case_reg}. 

\subsection{Regression of Bot Following Across Demographic and Personality Differences}
\label{regress_botfol}
We use a multiple linear regression model using Strata to model demography and personality-based differences in whether an individual follows the bot. Our results suggest that older individuals (p = 0.001) are less likely to follow the bot. The rest were not statistically significant. All results are shown in Table \ref{tab:multilinear_reg}. 

\subsection{Proportion of Bot Following Across Demographic and Personality Differences}
\label{prop_botfol}
\begin{figure}
    \centering
    \includegraphics[width=0.7\linewidth]{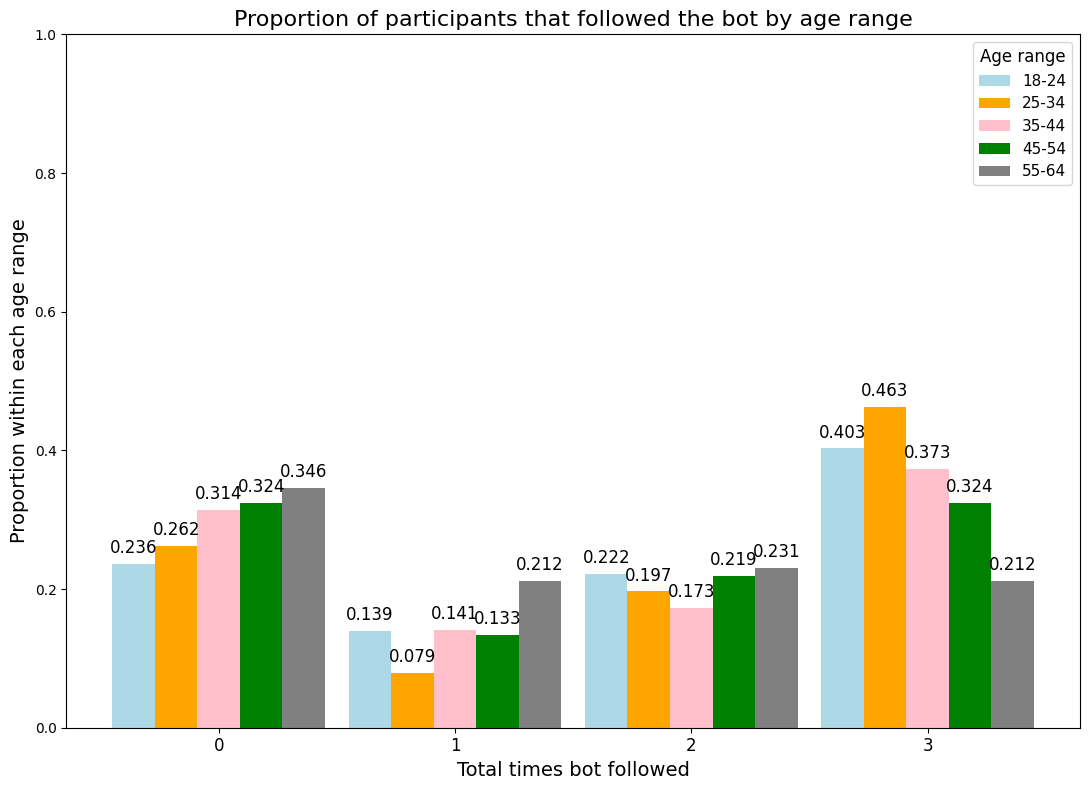}
    \caption{Proportion of participants and number of times they followed the bot by age range. There is no statistical difference based on age range. }
    \label{fig:follow_age}
\end{figure}

\begin{figure}
    \centering
    \includegraphics[width=0.7\linewidth]{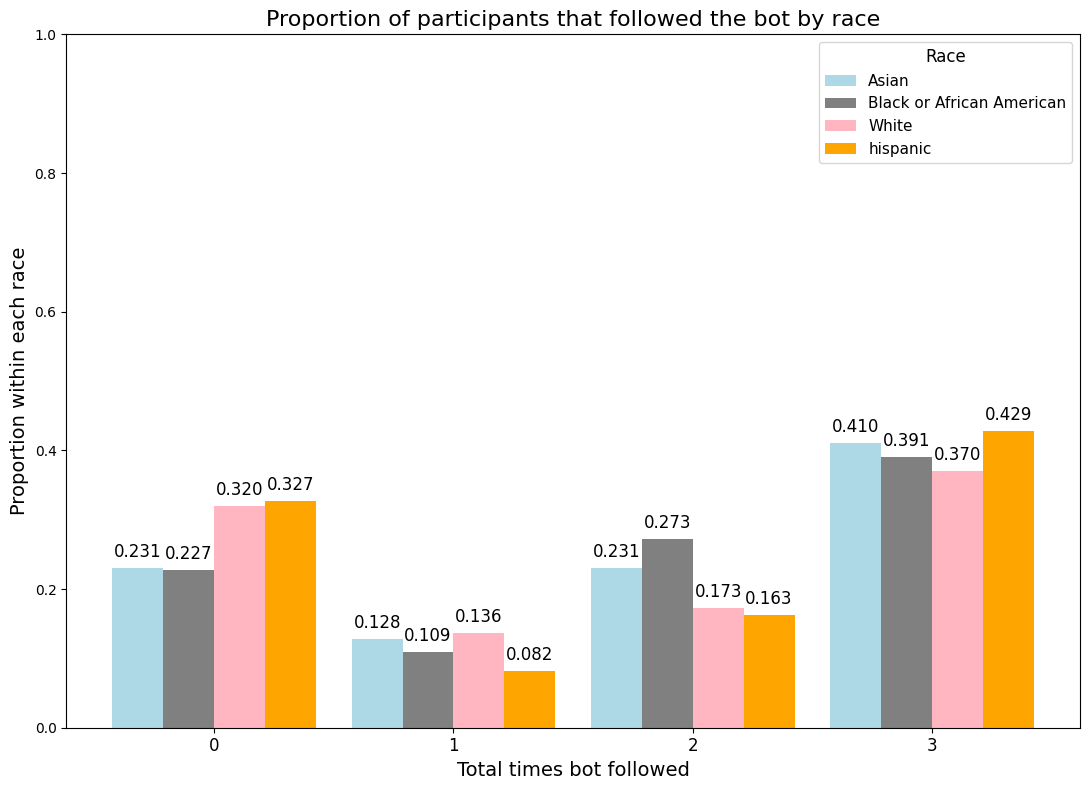}
    \caption{Proportion of participants and number of times they followed the bot by race. There is no statistical difference based on race. }
    \label{fig:follow_race}
\end{figure}

\begin{figure}
    \centering
    \includegraphics[width=0.7\linewidth]{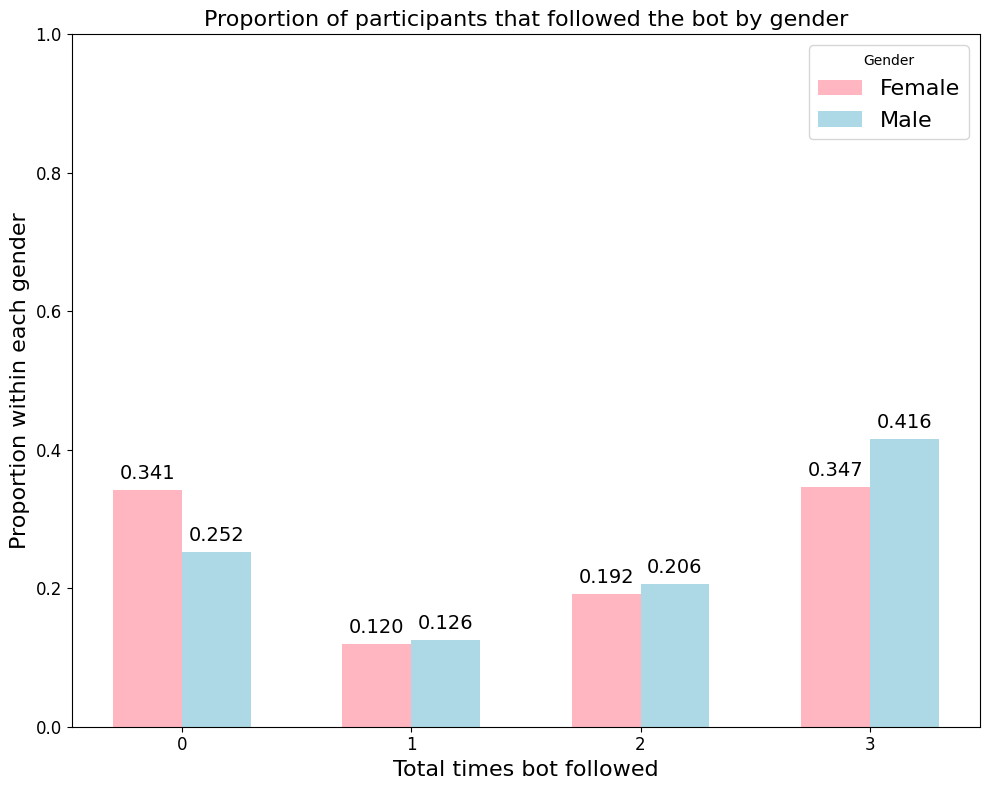}
    \caption{Proportion of participants and number of times they followed the bot by gender. There is no statistical difference based on gender. }
    \label{fig:follow_gender}
\end{figure}

\begin{figure}
    \centering
    \includegraphics[width=0.7\linewidth]{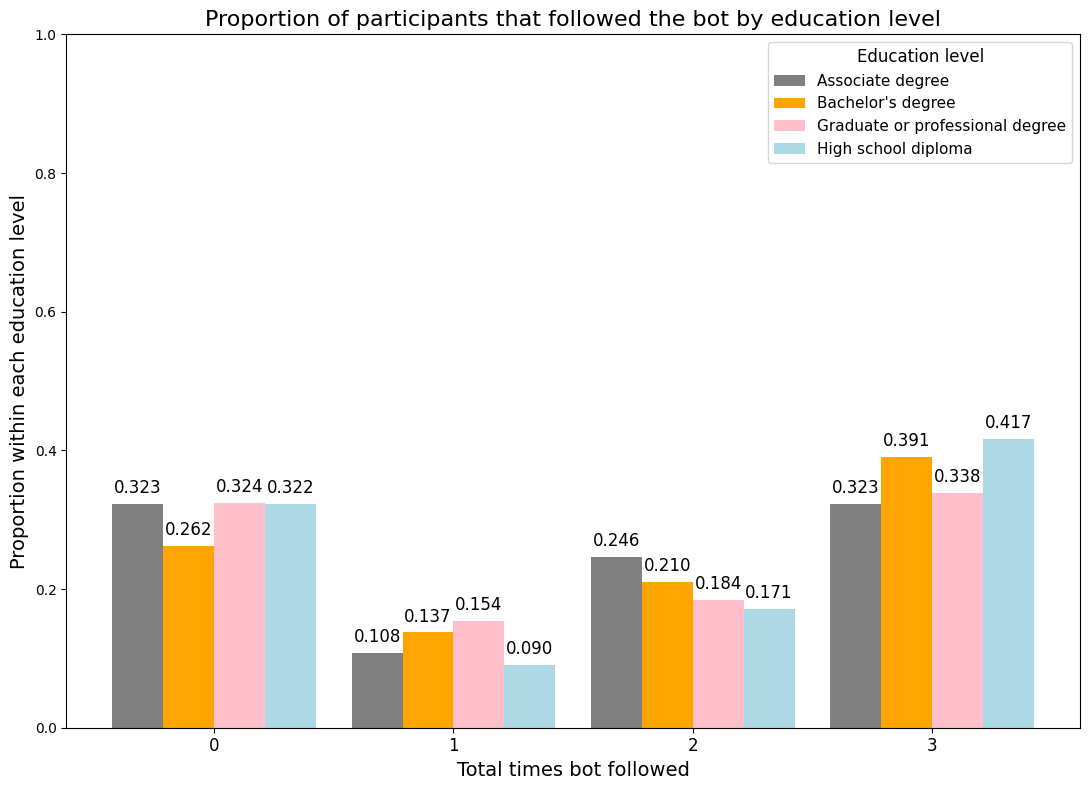}
    \caption{Proportion of participants and number of times they followed the bot by education. There is no statistical difference based on education levels. }
    \label{fig:follow_education}
\end{figure}

\begin{figure}
    \centering
    \includegraphics[width=0.7\linewidth]{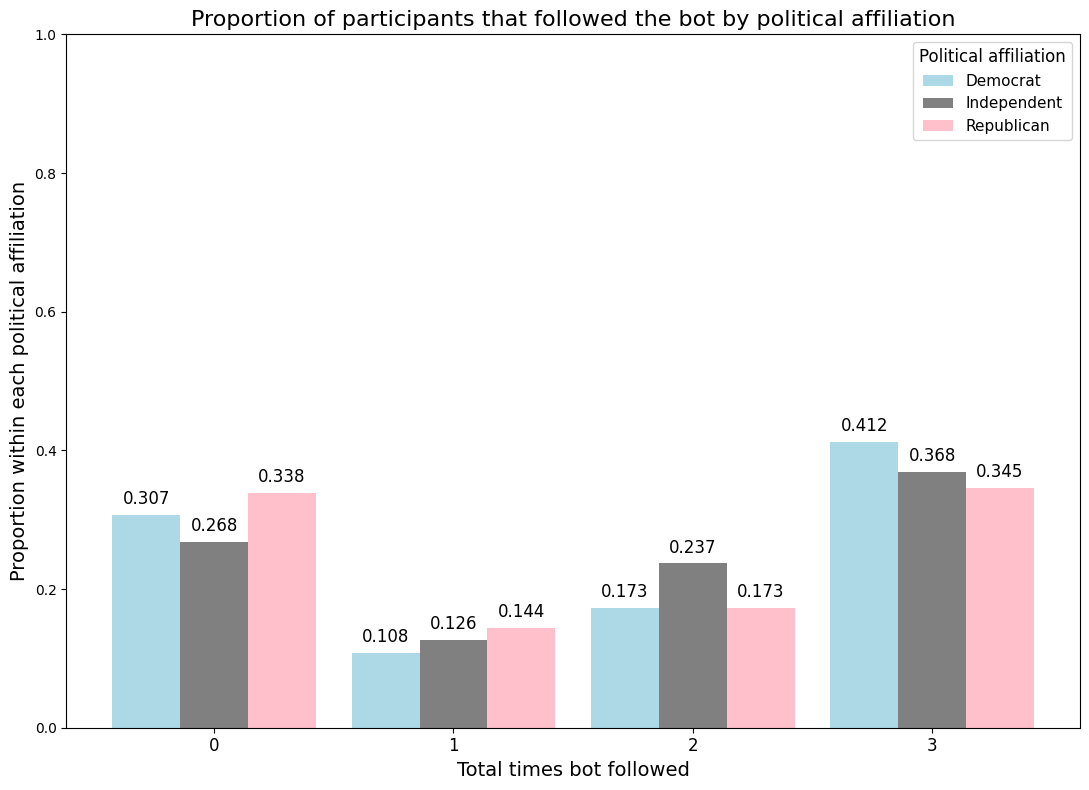}
    \caption{Proportion of participants and number of times they followed the bot by political affiliation. There is no statistical difference based on political affiliation. }
    \label{fig:follow_political}
\end{figure}

\begin{figure}
    \centering
    \includegraphics[width=0.7\linewidth]{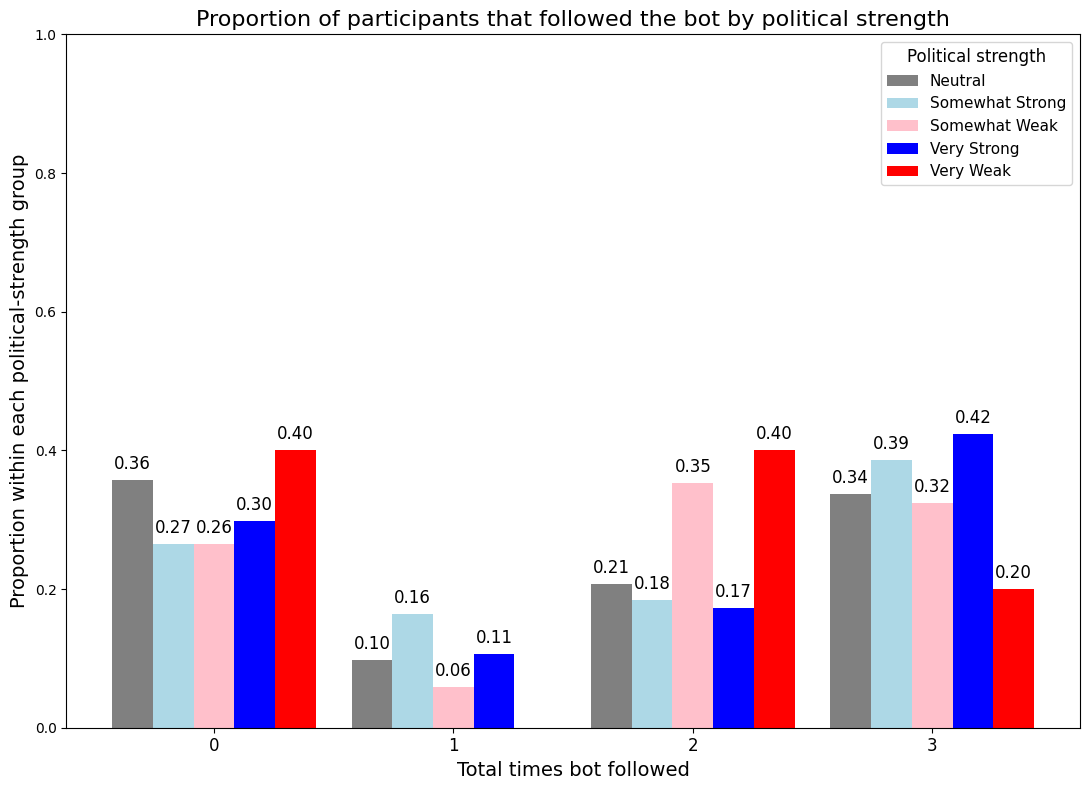}
    \caption{Proportion of participants and number of times they followed the bot by political strength. There is no statistical difference based on political strength.}
    \label{fig:follow_pol_strength}
\end{figure}

We conduct chi-square tests for demographics to see if there is any connection between who individuals follow and their demographics, complementing the regression model. We find that there is no significance difference based on age range (Chi-square = 20.426, p-value = 0.0594, dof = 12), education level (Chi-square = 9.200, p-value = 0.4190, dof = 9), gender (Chi-square = 6.228, p-value = 0.1010, dof = 3), race (Chi-square = 10.287, p-value = 0.3277, dof = 9), political affiliation (Chi-square = 6.513, p-value = 0.3682, dof = 6), and political strength (Chi-square = 18.248, p-value = 0.1084, dof = 12). The plots are in Figures \ref{fig:follow_age}, \ref{fig:follow_education}, \ref{fig:follow_gender}, \ref{fig:follow_race}, \ref{fig:follow_political}, \ref{fig:follow_pol_strength}.

\subsection{Pre-Election and Post-Election Checks of Differences}
\label{prepost}

\begin{figure}[htbp]
    \centering
    \includegraphics[width=0.5\linewidth]{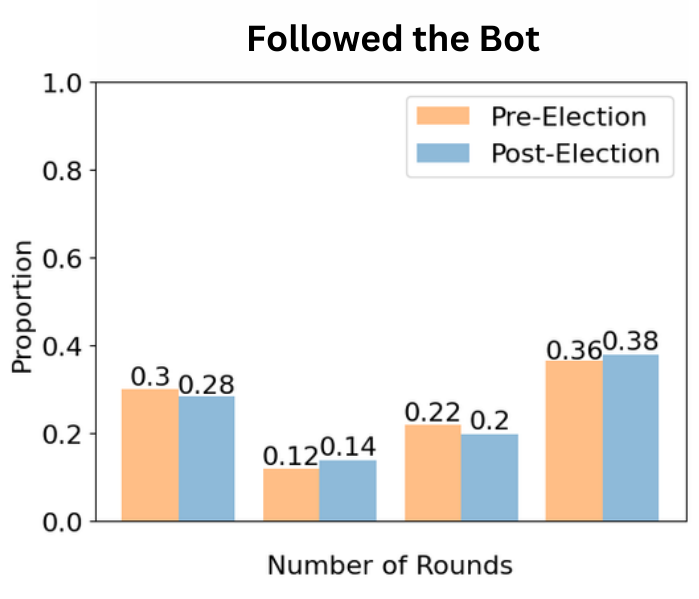}
    \caption{\textbf{Comparison of Treatment Proportions Pre-Election and Post-Election, Bot Following by Number of Rounds}. Distribution differences are insignificant (see Table \ref{tab:caseproofpropinsig} for specifics.)}
    \label{fig:bot_follow}
\end{figure}

\begin{figure}[htbp]
    \centering
    \includegraphics[width=1\linewidth]{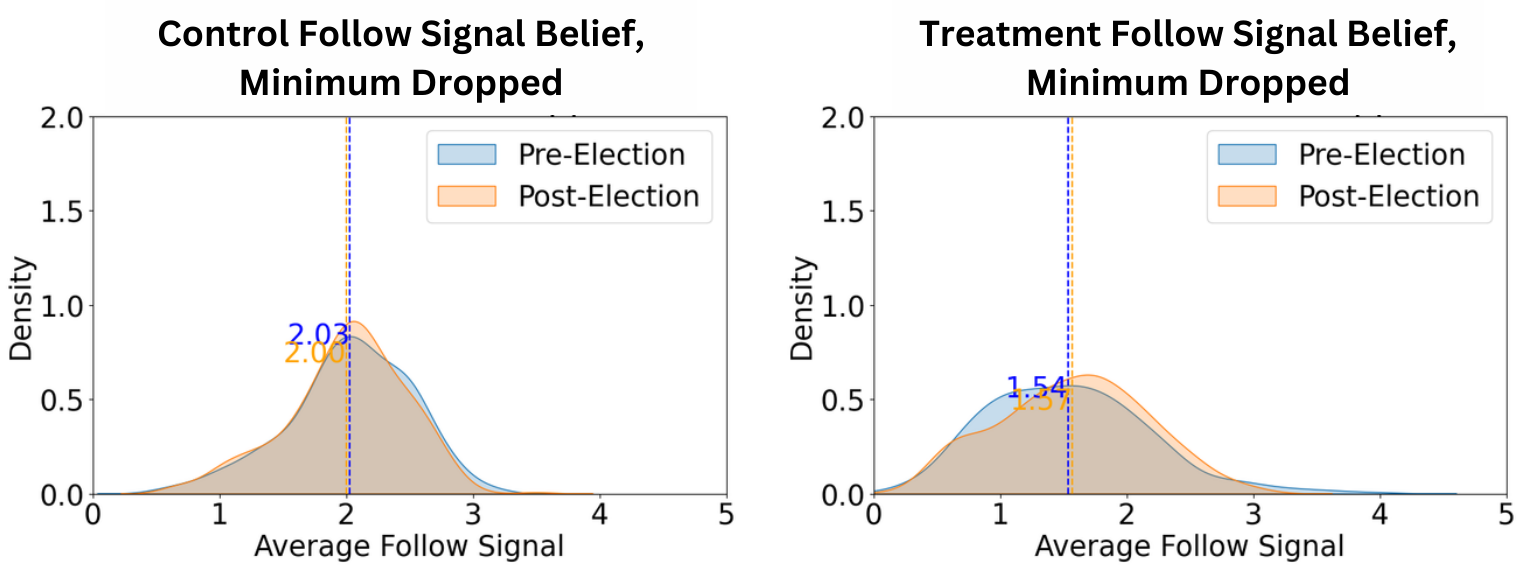}
    \caption{\textbf{Comparison of Follow Signal Distributions Pre-Election and Post-Election, Based on Treatment and Control}. Distribution differences are insignificant (see Table \ref{tab:caseproofofinsig} for specifics.)}
    \label{fig:followsignal}
\end{figure}

\begin{figure}[htbp]
    \centering
    \includegraphics[width=1\linewidth]{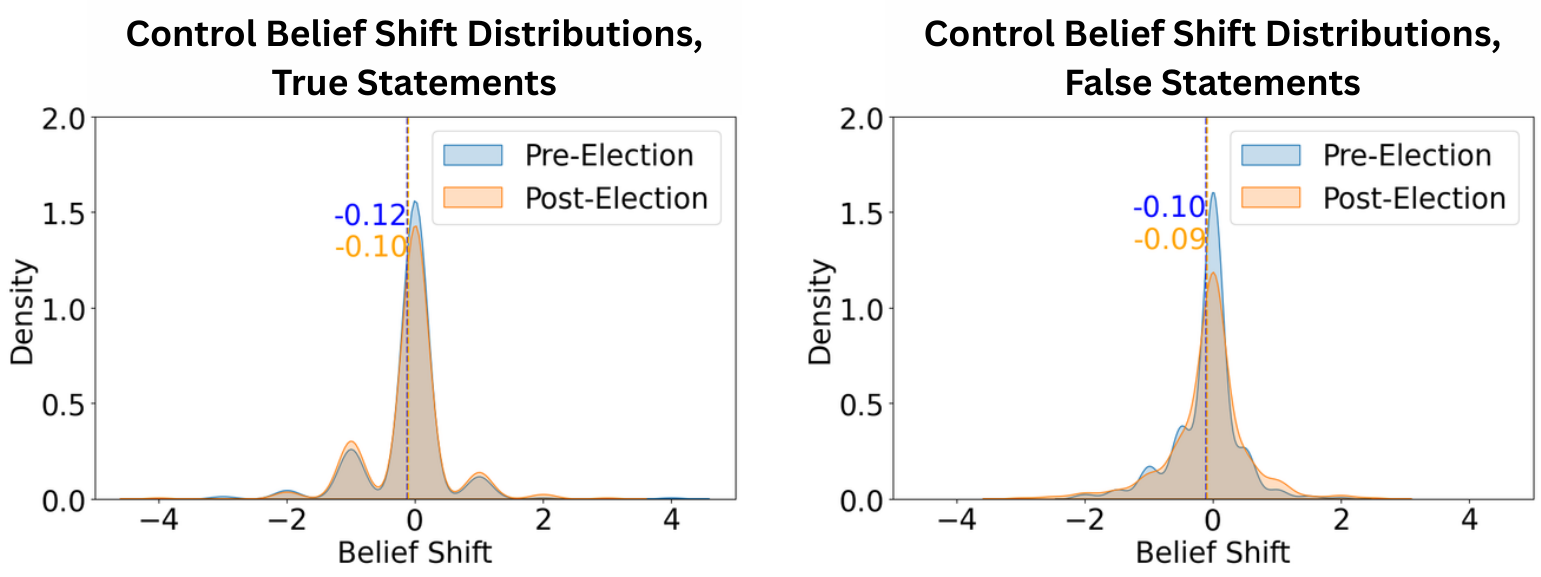}
    \caption{\textbf{Comparison of Control Distributions Pre and Post Election, Split Upon Ground Truth}. Distribution differences are insignificant (see Table \ref{tab:controlproofofinsig} for specifics.)}
    \label{fig:cont_groundtruth}
\end{figure}

\begin{figure}[htbp]
    \centering
    \includegraphics[width=1\linewidth]{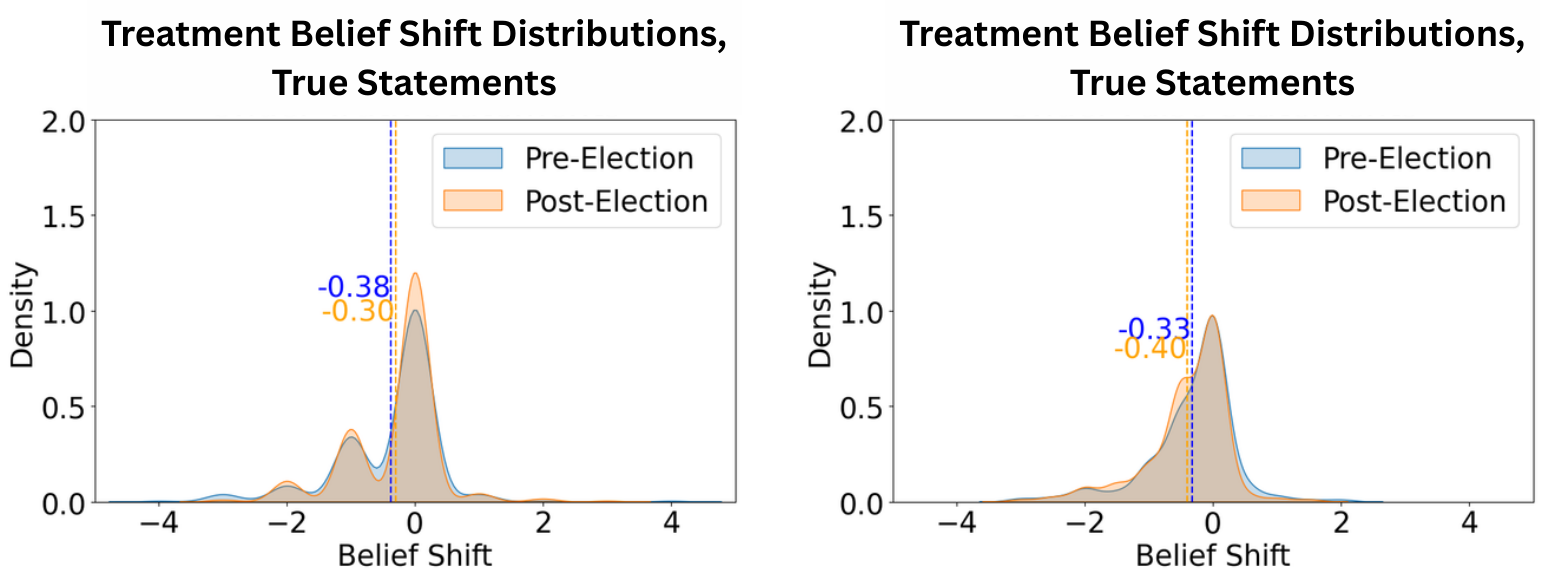}
    \caption{\textbf{Comparison of Treatment Distributions Pre-Election and Post-Election, Split Upon Ground Truth}. Distribution differences are insignificant (see Table \ref{tab:caseproofofinsig} for specifics.)}
    \label{fig:case_groundtruth}
\end{figure}

\begin{figure}[htbp]
    \centering
    \includegraphics[width=1\linewidth]{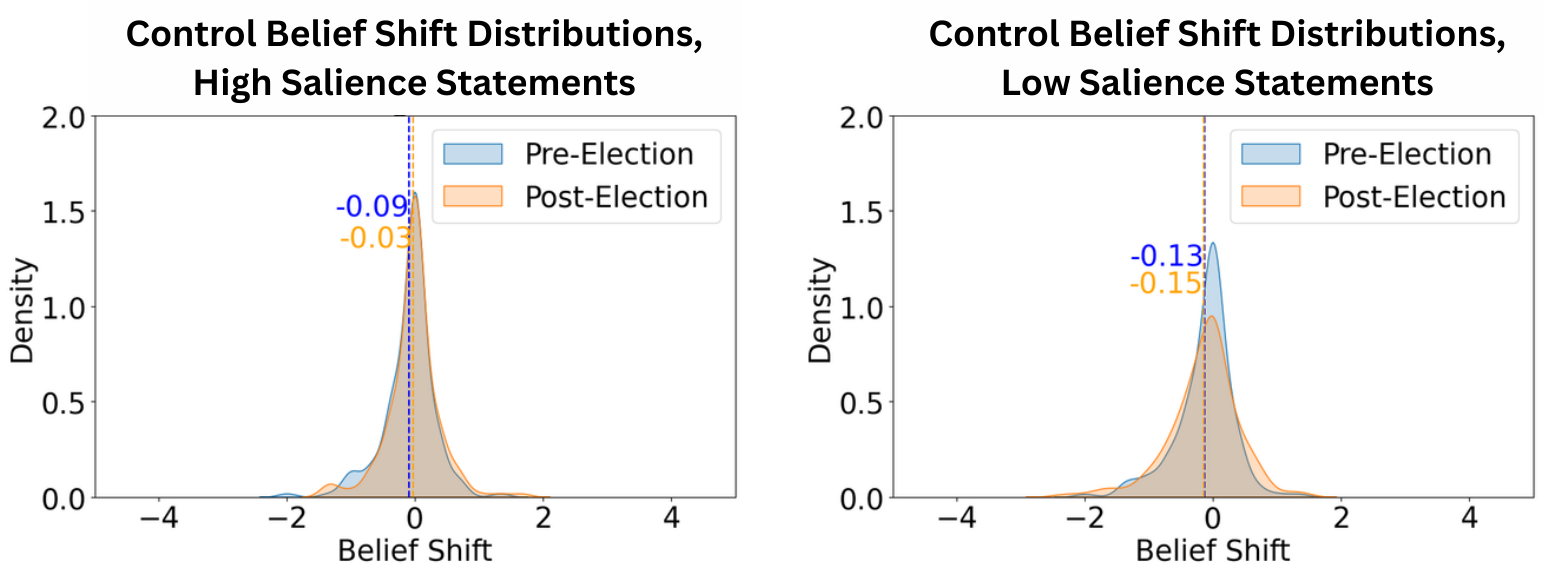}
    \caption{\textbf{Comparison of Control Distributions Pre-Election and Post-Election, Split Upon Salience of the Topics}. Distribution differences are insignificant (see Table \ref{tab:controlproofofinsig} for specifics.)}
    \label{fig:cont_salience}
\end{figure}

\begin{figure}[htbp]
    \centering
    \includegraphics[width=1\linewidth]{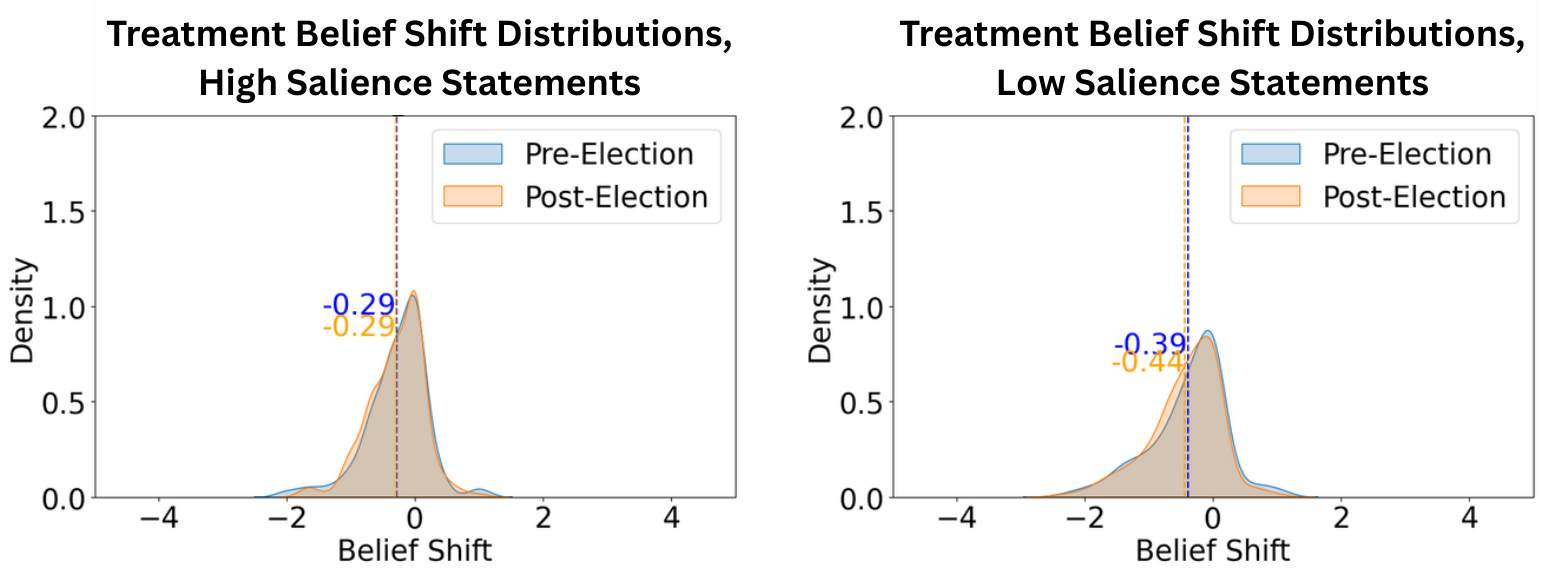}
    \caption{\textbf{Comparison of Treatment Distributions Pre-Election and Post-Election, Split Upon Salience of the Topics}. Distribution differences are insignificant (see Table \ref{tab:caseproofofinsig} for specifics.)}
    \label{fig:case_salience}
\end{figure}

Because data collection occurred before and after the election, we tested to identify differences in results between the treatment and control. We find no evidence of significantly diverging findings between the two periods (as shown in Supplemental figures). Thus, both samples were combined for the main paper results. The results splitting ``Pre-Election'' and ``Post-Election'',  are provided in Supplemental Figures \ref{fig:alldist_supplementals}, \ref{fig:cont_proportions}, \ref{fig:case_proportions}, \ref{fig:bot_follow}, \ref{fig:followsignal}, \ref{fig:cont_groundtruth}, \ref{fig:case_groundtruth}, \ref{fig:cont_salience}, \ref{fig:case_salience}, \ref{fig:cont_econ}, \ref{fig:case_econ}, \ref{fig:cont_poltrust}, and \ref{fig:case_poltrust}, and in Supplemental Tables \ref{tab:controlpvp}, \ref{tab:caseproofofinsig}, \ref{tab:casepvp}, \ref{tab:controlproofofinsig}, \ref{tab:caseproofofinsig}, \ref{tab:controlproofpropinsig}, and \ref{tab:caseproofpropinsig}.

\subsection{Study-Level Differences}
\label{studylevel}
\begin{figure}
    \centering
    \includegraphics[width=0.95\linewidth]{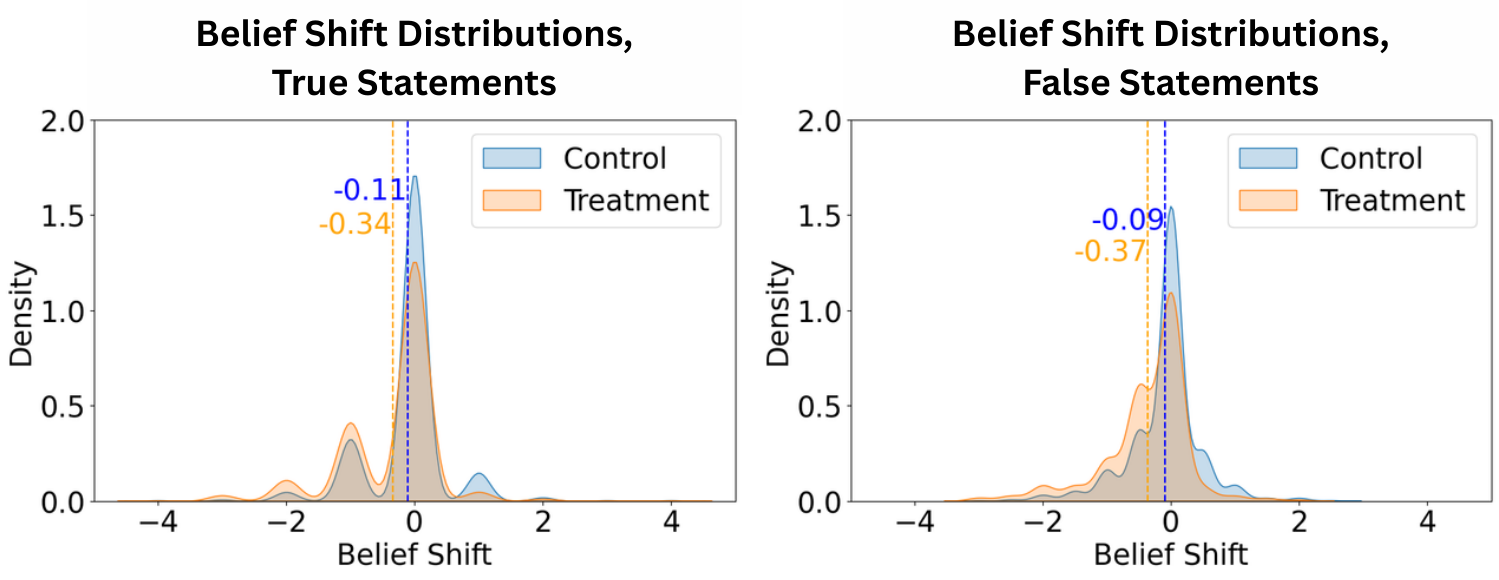}
    \caption{\textbf{Average shift of all participants, splitting into rounds with ``True'' ground truth and rounds with ``False'' ground truth.} The treatment distribution mean is more negative, implying individuals in the treatment shifted towards the truth across all of these conditions, and the result is significant.}
    \label{fig:finding3_split_truevfalse}
\end{figure}

\begin{figure}
    \centering
    \includegraphics[width=0.95\linewidth]{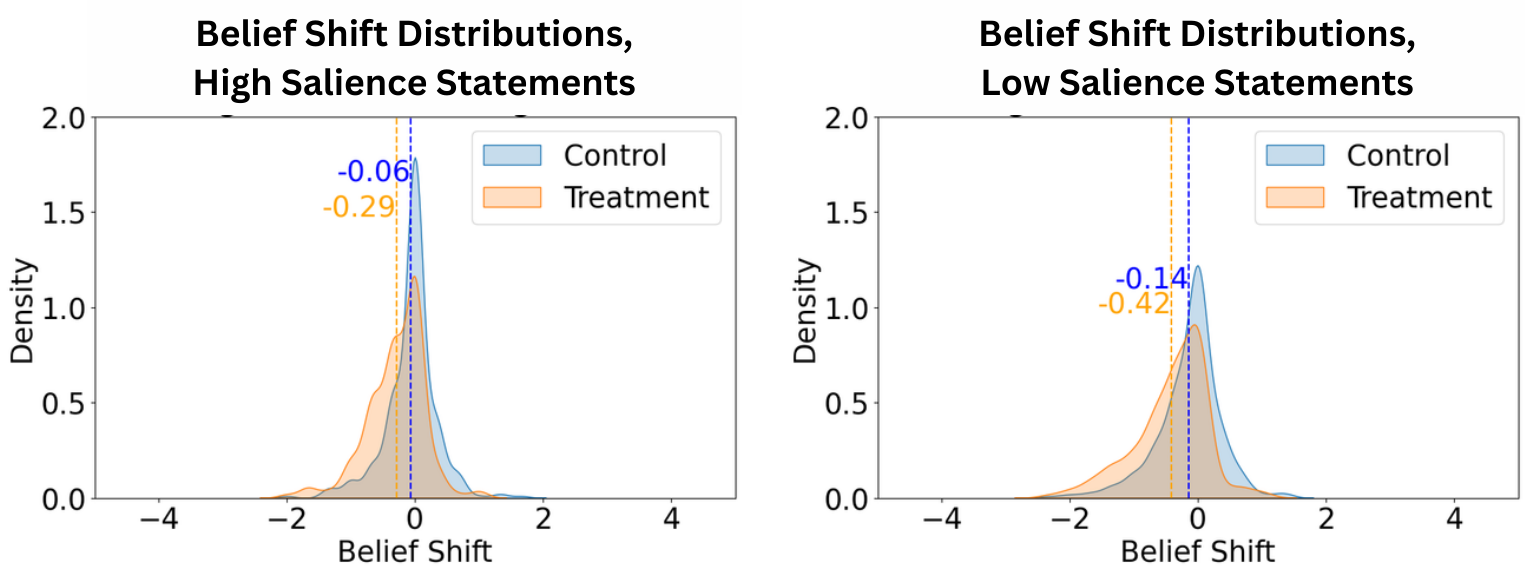}
    \caption{\textbf{Average shift of all participants, splitting by games that contained ``High Salience'' topics and ``Low Salience'' topics.} The treatment distribution mean is more negative, implying individuals in the treatment shifted towards the truth across all of these conditions, and the result is significant.}
    \label{fig:finding3_split_salience}
\end{figure}

\begin{figure}
    \centering
    \includegraphics[width=0.95\linewidth]{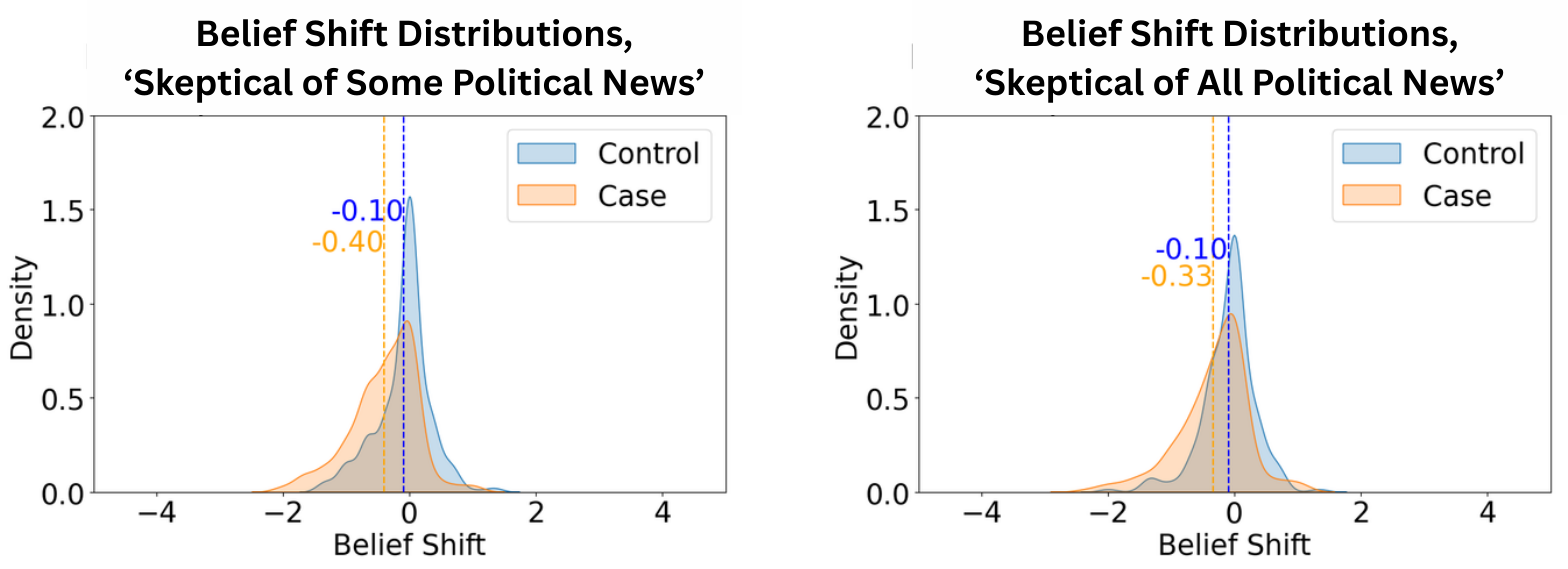}
    \caption{\textbf{Average shift of all participants, splitting by participants who self selected themselves to be ``Skeptical of some political news'' and ``Skeptical of All Political News''.} The treatment distribution mean is more negative, implying individuals in the treatment shifted towards the truth across all of these conditions, and the result is significant for the options of ``Skeptical of Some Political News'' and ``Skeptical of All Political News''. 0 participants chose they    ``Trusted All Political News'' and very few participants selected that they ``Trusted Some Political News'', leading to insignificant results in this category.}
    \label{fig:finding3_split_newstrust}
\end{figure}

We empirically evaluate the efficacy of the intervention across the veracity of the statements (true/false), saliency levels of the topic, economic relevancy of the topic, and the reported skepticism in political news (see Figures~\ref{fig:finding3_split_truevfalse}, \ref{fig:finding3_split_salience}, \ref{fig:finding3_split_salience}, and \ref{fig:finding3_split_newstrust} and Table \ref{tab:robustness}). We find that across all measures, the treatment significantly moves closer to the truth than the control. The only exception to this phenomenon was for responses to ``Trust all political news'' (which 0 participants selected) and ``Trust some political news'' (which few participants selected, leading to poor statistical power), where the difference was insignificant. 

\subsection{Self-Reported Belief Updates}
\label{selfreportbeliefupdate}
\begin{figure}
    \centering
    \includegraphics[width=0.75\linewidth]{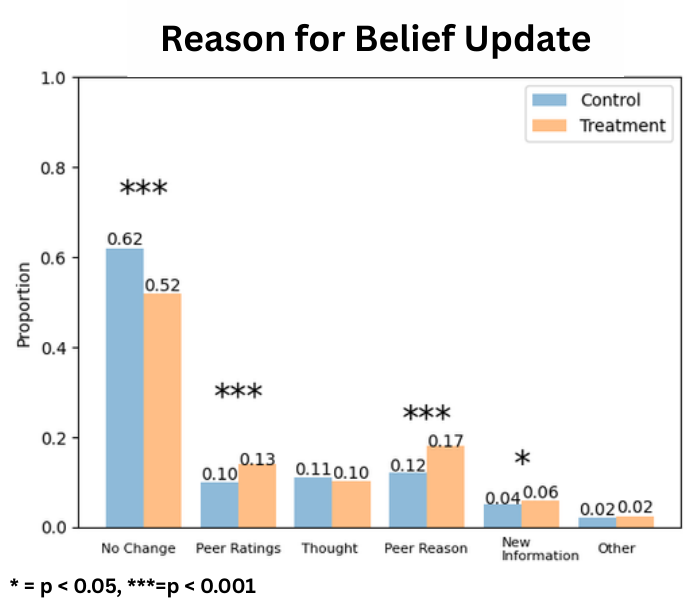}
    \caption{\textbf{Proportion of Responses for Belief Update in Stage 2, Across Treatment and Control.} `***' represents significance at p \textless 0.001, `***' represents significance at p \textless 0.01, and `*' represents significance at p \textless 0.05}
    \label{fig:stage2_beliefupdatereason}
\end{figure}

Within the experiments, at stage 2, participants are asked to select reasons as to why they are updating their opinion, which include options such as the following:

\begin{itemize}
    \item ``I did not change my mind'' (abbreviated to ``No change'' in Figure \ref{fig:stage2_beliefupdatereason})
    \item ``The believability ratings of my peers changed my mind'' (abbreviated to ``Peer ratings'' in Figure \ref{fig:stage2_beliefupdatereason})
    \item ``I thought about the topic more on my own'' (abbreviated to ``Thought'' in Figure \ref{fig:stage2_beliefupdatereason})
    \item ``The reasons stated in the answer of my peers changed my mind'' (abbreviated to ``Peer reason'' in Figure \ref{fig:stage2_beliefupdatereason})
    \item ``I found new information from other sources not related to this experiment (eg. social media, news, friends and family)'' (abbreviated to ``New information'' in Figure \ref{fig:stage2_beliefupdatereason})
    \item ``Other''
\end{itemize}

From Figure \ref{fig:stage2_beliefupdatereason}, we find that individuals in the treatment selected ``No change'' less frequently than the control, and ``Peer ratings'', ``peer reasoning'', and ``new information'' more often than the control. We attribute the significance of the ``Peer ratings'' and ``peer reasoning'' to the presence of the bot, as this is the only difference between the treatment and control in stage 2. While it is unclear what ``new information'' could refer to, it could also be interpreted as individuals feeling moved by the facts and sources cited by the bot.

\subsection{The connection between changing rating and following the bot}
\label{botfolvchange}
To understand if there is a connection between individual rating and following the bot, we take a more granular look at the treatment group. For each round for each participant, we consider if they followed the bot or not, and calculate their belief change (with a more negative rating meaning moving closer to the truth). Our results show that the mean change in rating was almost twice as high when they followed the bot compared to when they did not follow the bot. Mean change was -0.473 (CI: -0.526, -0.419) for when they followed the bot, and mean was -0.210 (CI: -0.261, -0.159), when they did not follow the bot. This is consistent with the possibility that the individuals altered their beliefs due to the bot, and then chose to follow the bot. We cannot, however, directly test that dynamic since we cannot isolate the precise actor in the network that influenced a respondent’s beliefs.

\subsection{Network-Level Following Trends}
\label{networklevel}
Given the main paper focuses on effects at the individual level, we also look at network effects. First, we examine the assortativity of the network, or how often individuals connect with others of similar beliefs. We find that individuals move towards homophily. This analysis is done by comparing stage 2 and stage 3, since individuals have a choice in stage 3 but not in stage 2 (all stage 2 and stage 3 differences significant at p \textless 0.001). We highlight these findings to show that individuals in both conditions are not updating their preferences at random. 

Since individuals in both the treatment and control are not updating their following at random, we then investigate how often individuals follow the bot per round, compared with how often individuals follow the ``most truthful'' option in control. When there is potentially more than one ``most truthful'' option (for example, two people have the same value), we select one from the set at random and check whether the individual followed them. This is meant to be comparable to the bot follow in the treatment, since the bot can be considered as only one ``user". Here we see similar trends in following these individuals between the treatment and the control (p = 0.118), suggesting that individuals in both the treatment and the control tend to pursue others with more truthful beliefs.

% We also look into the assortativity of the network, or how often individuals connect with others of similar beliefs. We find that individuals move towards homophily. This analysis is done by comparing stage 2 and stage 3, since individuals have a choice in stage 3 but not in stage 2 (all stage 2 and stage 3 differences significant at p \textless 0.001). We highlight these findings to show that individuals in both conditions are not updating their preferences at random, but rather are updating their networks towards the truth, and argue this pursuit of truth can be leveraged for supporting LLM interventions such as ours in the future. 

\begin{figure}
    \centering
    \includegraphics[width=1\linewidth]{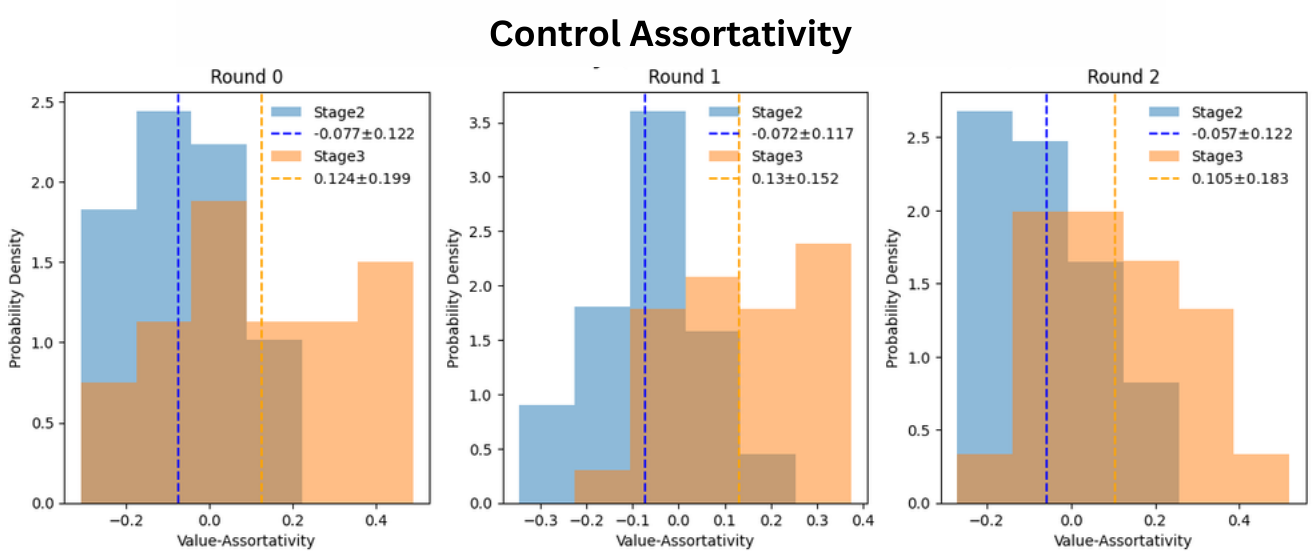}
    \caption{\textbf{Assortativity Measures in Control, Comparing Stage 2 and Stage 3.} All comparisons of stage 2 and stage 3 distributions are significant, with stage 3 shifting towards a more assortative network.}
    \label{fig:assortativity_control}
\end{figure}

\begin{figure}
    \centering
    \includegraphics[width=1\linewidth]{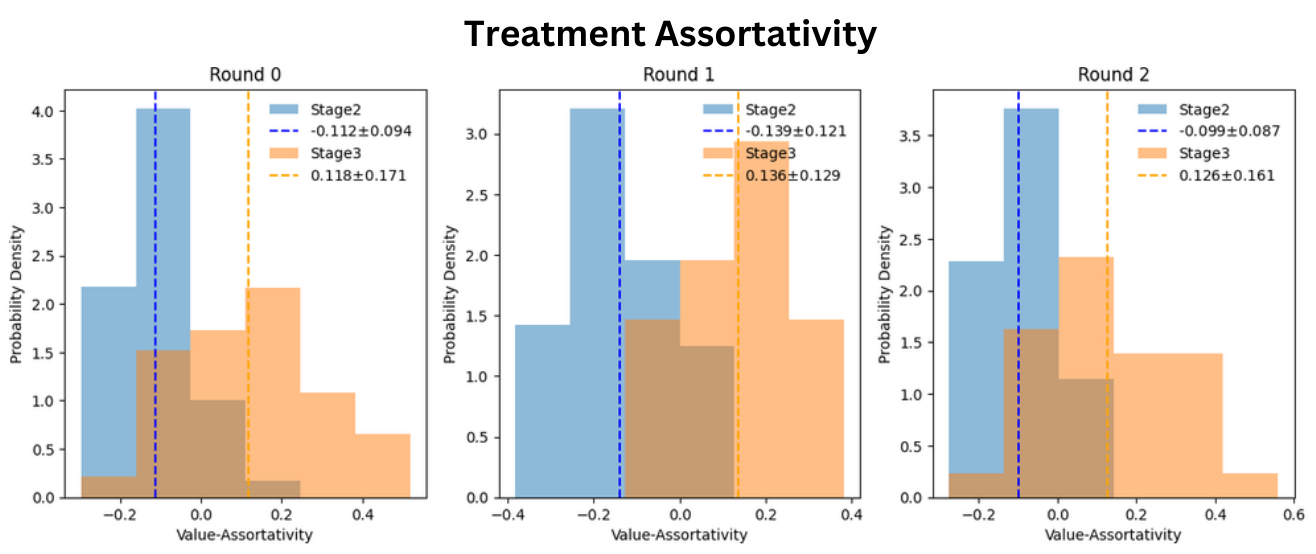}
    \caption{\textbf{Assortativity Measures in Treatment, Comparing Stage 2 and Stage 3.} All comparisons of stage 2 and stage 3 distributions are significant, with stage 3 shifting towards a more assortative network.}
    \label{fig:assortativity_case}
\end{figure}

\begin{figure}[ht]
    \centering
    \includegraphics[width=1\linewidth]{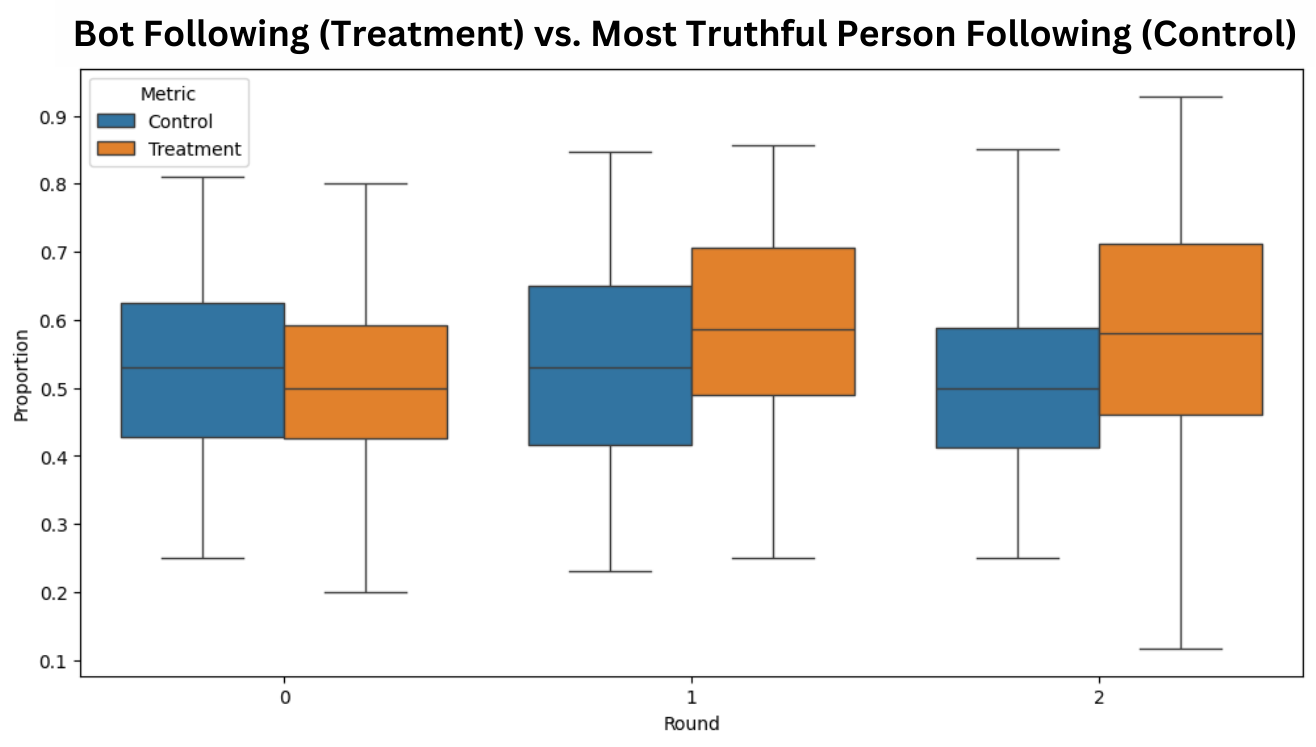}
    \caption{\textbf{Proportion of Individuals Following the Bot per Round Compared to Individuals Following a Randomly Selected ``Most Truthful'' Option in Control.} The trends are similar between the control and the treatment, suggesting truthful networks are pursued in both conditions.}
    \label{fig:botfollowvrandomtruth}
\end{figure}

These findings in tandem show that the individuals within the experiment are utilizing the follow mechanic as intended, and individuals pursue truth within their networks. The plots are in Figures \ref{fig:botfollowvrandomtruth}, \ref{fig:assortativity_control} and \ref{fig:assortativity_case}. We argue this pursuit of truth can be leveraged for supporting LLM interventions such as ours in the future. 

\subsection{Persuasiveness of the LLM Bot}
\label{persuasivellm}
All previous work discussed in this manuscript points to the implication that our LLM intervention is, in fact, persuasive. Prior works have emphasized that LLMs can embody persuasive traits \cite{costello2024durably, breum2024persuasive, Argyle_Busby_Gubler_Lyman_Olcott_Pond_Wingate_2025}. That being said, we are interested in understanding why our LLM intervention is persuasive. In order to identify what aspects of the intervention make it persuasive, we analyzed message properties from Stage 1 for both the participants and the LLM intervention, comparing the shifts. We find that personal references within participant responses and displaying a mix of partisan cues through source references (compared to one partisan cue) were significant. 

\begin{figure}[ht]
    \centering
    \includegraphics[width=1\linewidth]{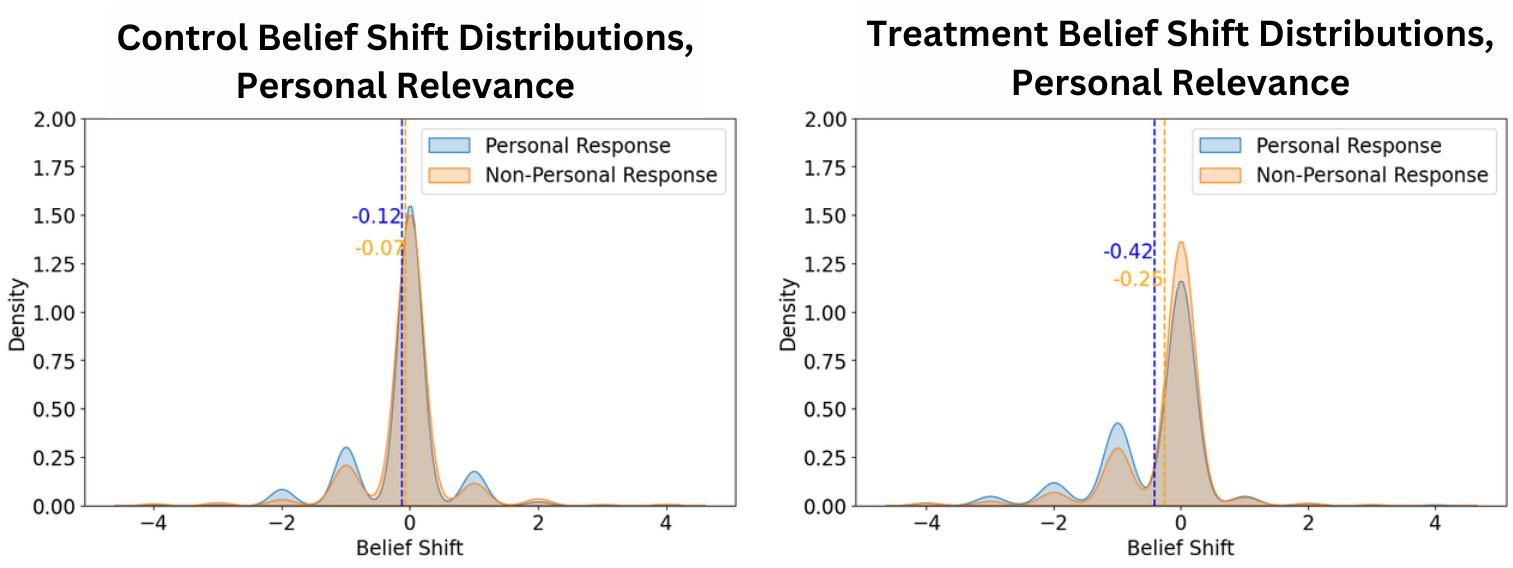}
    \caption{\textbf{Belief Shifts By Round, Partitioned on Responses that Utilize Personal Pronouns vs. No Personal Pronouns}. Control difference in distribution is not significant, while treatment difference is. Results show individuals who wrote responses that utilized personal pronouns (``I'', ``our'', ``we'') shifted more towards the truth.}
    \label{fig:botpersuasion_personal}
\end{figure}

Figure \ref{fig:botpersuasion_personal} shows the results of the control and treatment shifts, partitioned on whether or not the participant utilized personal pronouns within their Stage 1 written response. Personal pronouns include ``I'', ``we'', ``our'', ``us'', and variants of those including contractions. ``us'' was explicitly checked for whether or not it was referring to ``US'' (United States) since many of our statements are about the US at large. We find no notable difference in the control conditions (p = 0.179), while the treatment condition shows a larger shift when individuals reference personal pronouns (p \textless 0.00001). Given this is only significant in the treatment, this suggests that the bot has more of an influence on individuals who feel the topic is personally relevant, which aligns with prior persuasion work \cite{Luttrell_2018, Petty_Krosnick_2014}.

\begin{figure}
    \centering
    \includegraphics[width=1\linewidth]{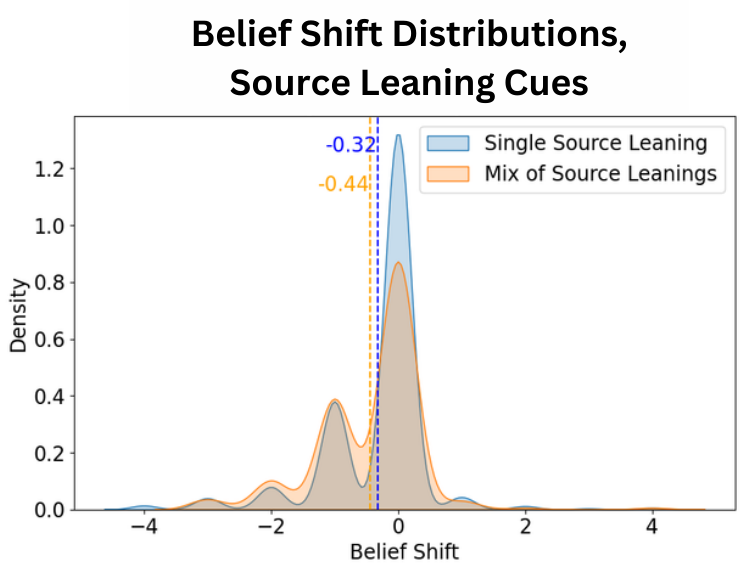}
    \caption{\textbf{Belief Shifts By Round, Partitioned on Whether the Bot Response Included a Mix of Partisan Cues or a Single Partisan Cue in Source References}. Results indicate a higher shift when the LLM intervention offered a mix of sources.}
    \label{fig:botpersuasion_sourcelean}
\end{figure}

Figure \ref{fig:botpersuasion_sourcelean} shows the results of the treatment shifts, partitioned on whether the LLM intervention pulled in sources of a single political leaning compared against when it showed sources of different political leanings. The results show that individuals shifted more towards the truth when a variety of source leanings is provided (p \textless 0.05), though this happens somewhat infrequently throughout the experiments (n = 241 where individuals saw a mix of political leanings in sources in a given round, n = 962 where individuals saw one political leaning cue in a given round). This suggests that pulling in cross-partisan agreement can boost persuasiveness by mitigating partisan motivated reasoning, as shown in prior works \cite{Bolsen_Druckman_Cook_2014, druckman2003framing}.

\clearpage
\newpage

\section*{Figures}

\clearpage
\newpage

% \subsection*{Experiment Framework Figures}

\begin{figure}[ht]
    \centering
    \includegraphics[width=0.95\linewidth]{supplemental_imgs/framework/stage1.pdf}
    \caption{\textbf{Stage 1 of the Experiment.} In stage 1, users are prompted with a statement with a pre-determined factuality label. Participants select on the provided scale their belief on the statement's factuality, and reasoning for their beliefs.}
    \label{fig:stage1}
\end{figure}

\begin{figure}[ht]
    \centering
    \includegraphics[width=0.65\linewidth]{supplemental_imgs/framework/stage2.pdf}
    \caption{\textbf{Stage 2 of the Experiment.} In stage 2, participants are shown a few statements from other participants (or the personalized bot, shown as Player 11). Participants then can update their beliefs and select a reason for the change.}
    \label{fig:stage2}
\end{figure}

\begin{figure}[ht]
    \centering
    \includegraphics[width=0.65\linewidth]{supplemental_imgs/framework/stage3.pdf}
    \caption{\textbf{Stage 3 of the Experiment.} In stage 3, participants are shown more statements from other participants (or the personalized bot, shown as Player 11). Participants can then update their networks, choosing to follow or unfollow the players shown.}
    \label{fig:stage3}
\end{figure}

% \subsection*{Pre-Election and Post-Election Checks of Difference}

\begin{figure}[ht]
    \centering
    \includegraphics[width=1\linewidth]{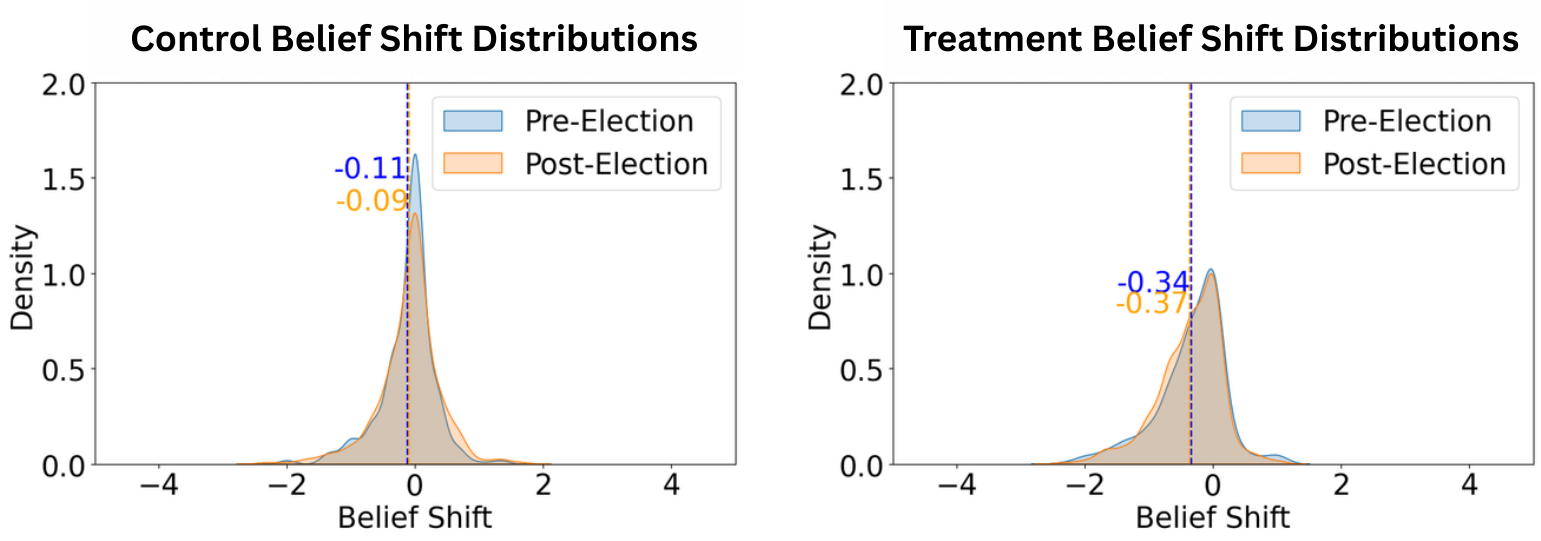}
    \caption{\textbf{Comparison of Distributions Pre-Election and Post-Election, Based on Treatment and Control}. Distribution differences are insignificant between pre-control and post-control, as well as pre-treatment and post-treatment (see Table \ref{tab:controlproofofinsig} and Table \ref{tab:caseproofofinsig} for specifics.)}
    \label{fig:alldist_supplementals}
\end{figure}

\begin{figure}[ht]
    \centering
    \includegraphics[width=1\linewidth]{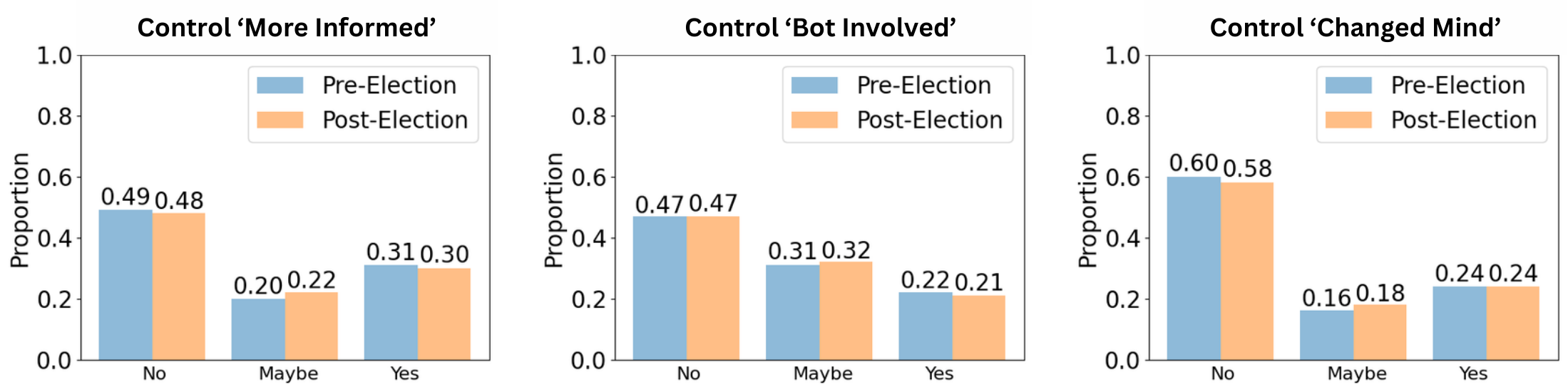}
    \caption{\textbf{Comparison of Control Proportions Pre-Election and Post-Election, Exit Survey Responses}. Distribution differences are insignificant (see Table \ref{tab:controlproofpropinsig} for specifics.)}
    \label{fig:cont_proportions}
\end{figure}

\begin{figure}[ht]
    \centering
    \includegraphics[width=1\linewidth]{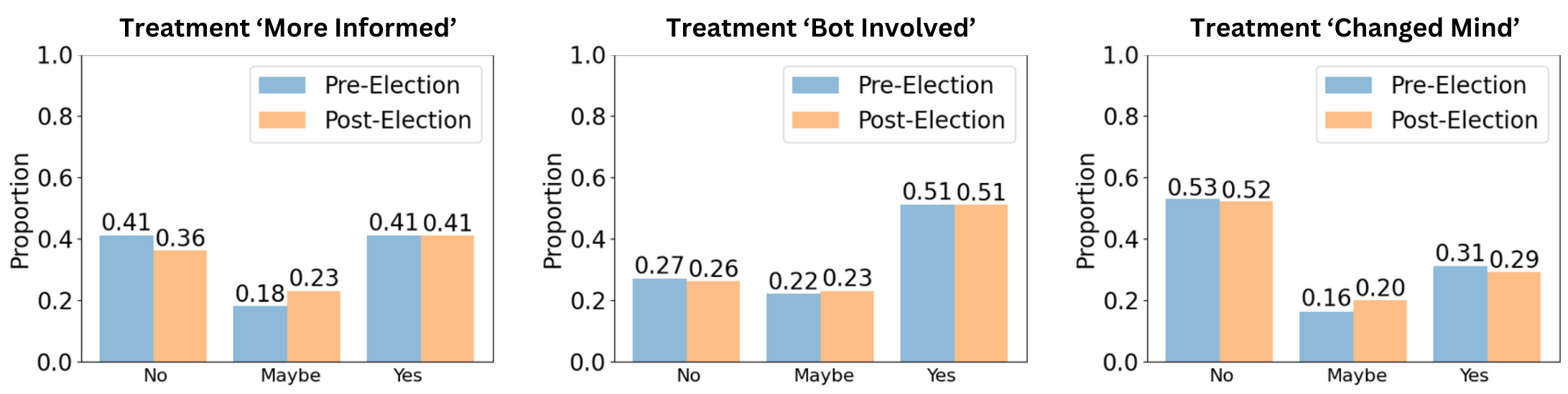}
    \caption{\textbf{Comparison of Treatment Proportions Pre-Election and Post-Election, Exit Survey Responses}. Distribution differences are insignificant (see Table \ref{tab:caseproofpropinsig} for specifics.)}
    \label{fig:case_proportions}
\end{figure}

\begin{figure}[htbp]
    \centering
    \includegraphics[width=1\linewidth]{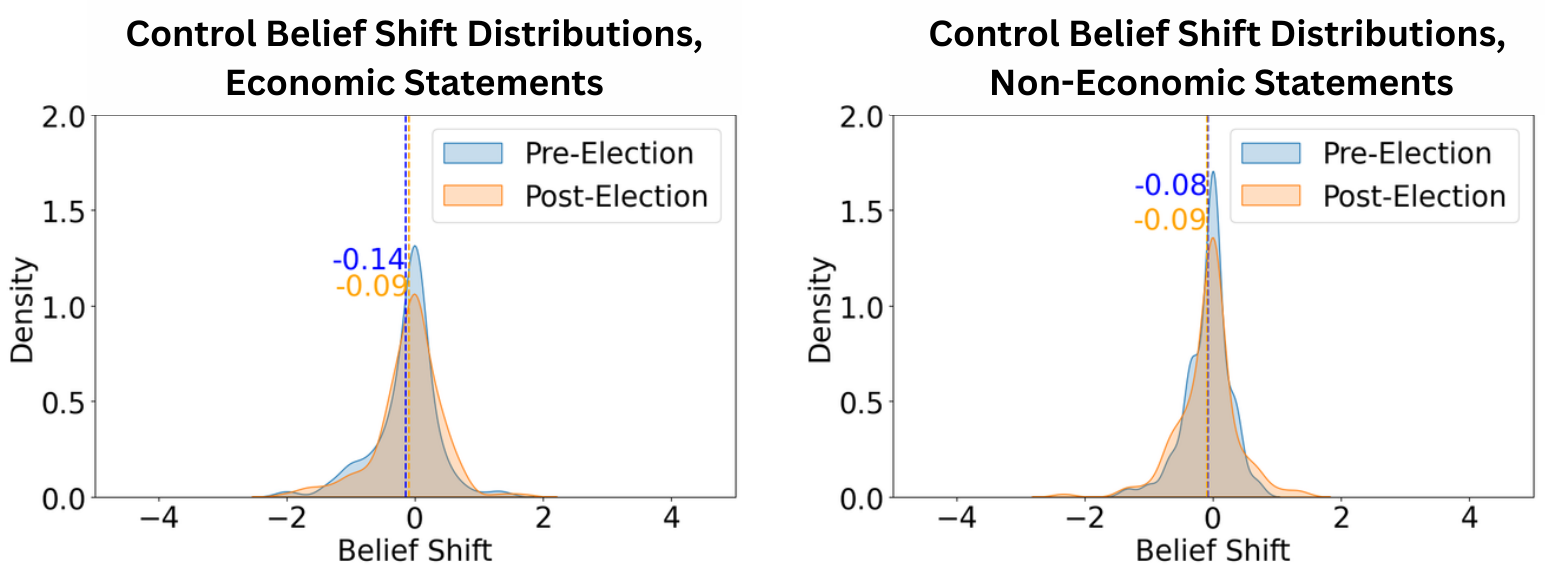}
    \caption{\textbf{Comparison of Control Distributions Pre-Election and Post-Election, Split Upon Economic Relevance of the Topic}. Distribution differences are insignificant (see Table \ref{tab:controlproofofinsig} for specifics.)}
    \label{fig:cont_econ}
\end{figure}

\begin{figure}[ht]
    \centering
    \includegraphics[width=1\linewidth]{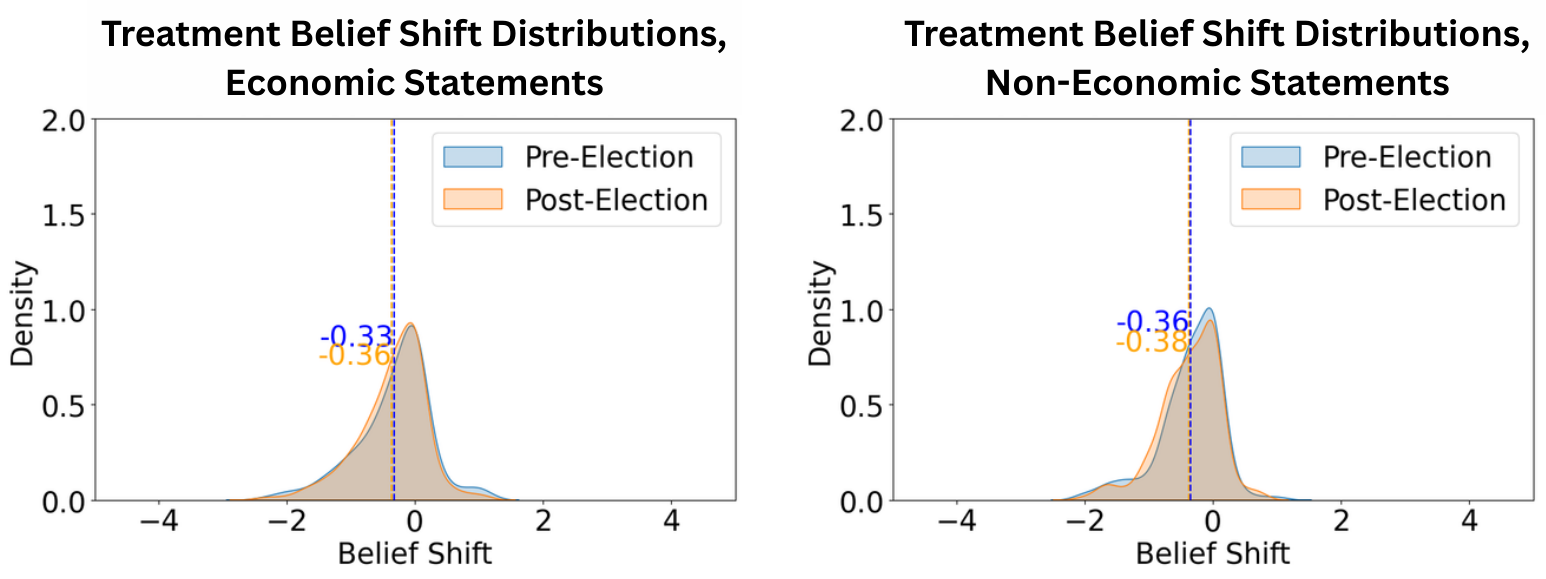}
    \caption{\textbf{Comparison of Treatment Distributions Pre-Election and Post-Election, Split Upon Economic Relevance of the Topic}. Distribution differences are insignificant (see Table \ref{tab:caseproofofinsig} for specifics.)}
    \label{fig:case_econ}
\end{figure}

\begin{figure}[ht]
    \centering
    \includegraphics[width=1\linewidth]{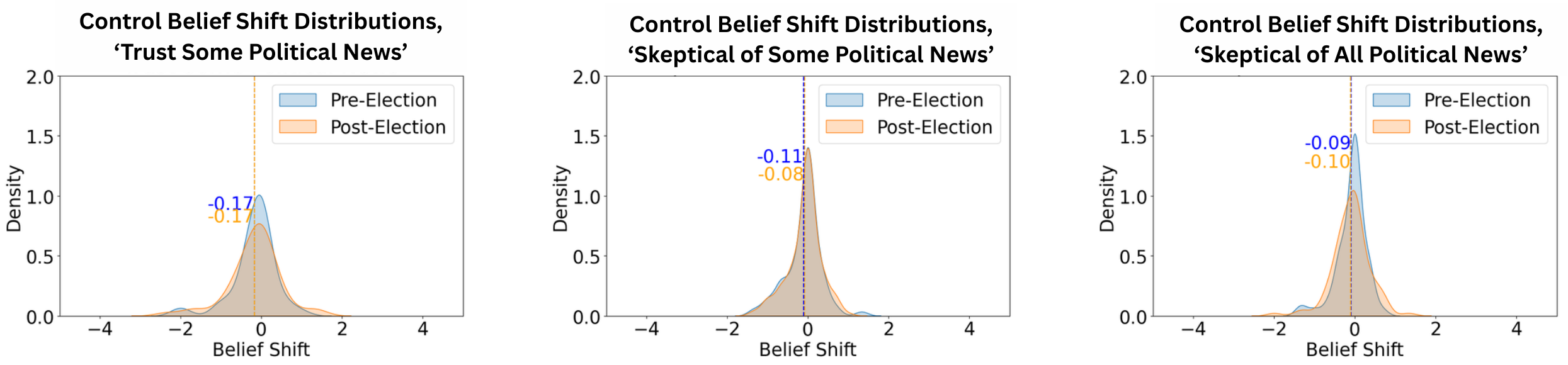}
    \caption{\textbf{Comparison of Control Distributions Pre-Election and Post-Election, Split Upon Self-Reported Political Trust}. Distribution differences are insignificant (see Table \ref{tab:controlproofofinsig} for specifics.)}
    \label{fig:cont_poltrust}
\end{figure}

\begin{figure}[ht]
    \centering
    \includegraphics[width=1\linewidth]{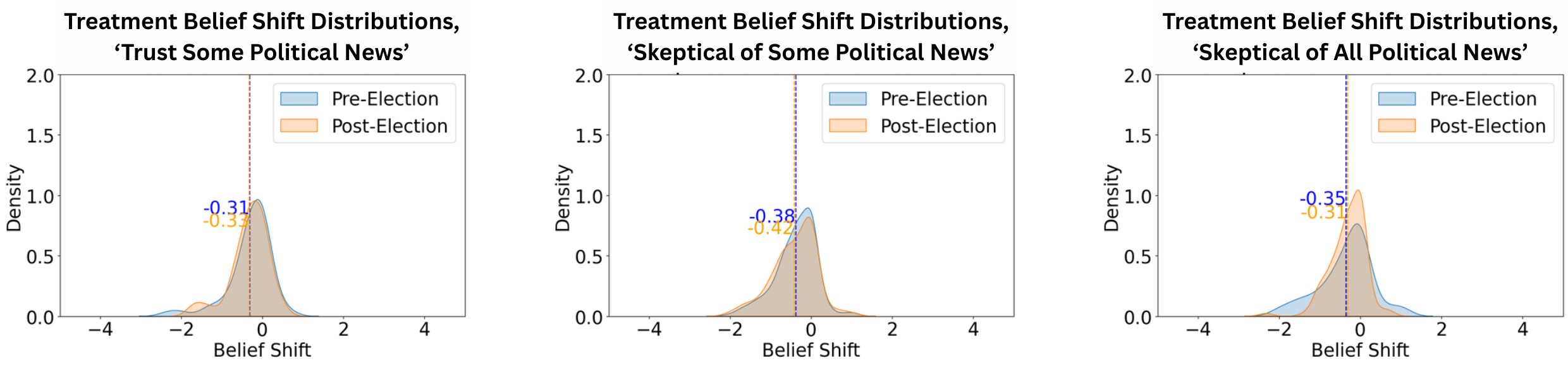}
    \caption{\textbf{Comparison of Treatment Distributions Pre-Election and Post-Election, Split Upon Self-Reported Political Trust}. Distribution differences are insignificant (see Table \ref{tab:caseproofofinsig} for specifics.)}
    \label{fig:case_poltrust}
\end{figure}

\begin{figure}
    \centering
    \includegraphics[width=0.95\linewidth]{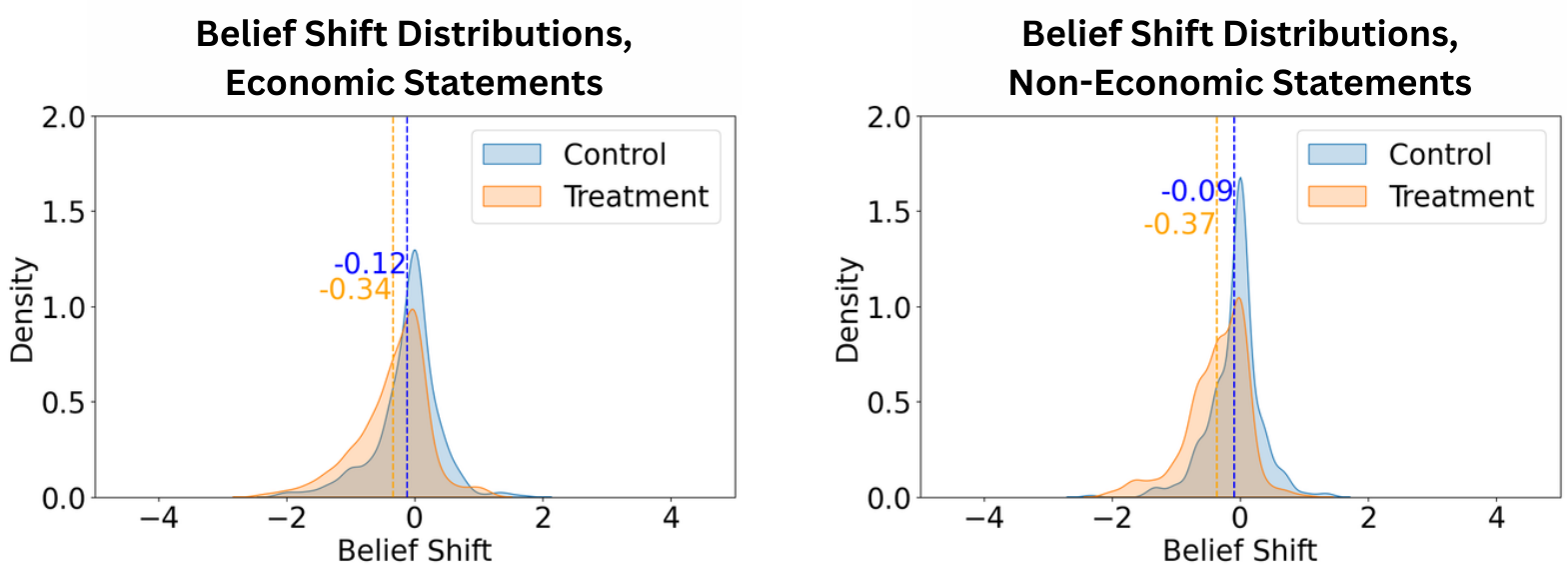}
    \caption{\textbf{Average shift of all participants, splitting by games that contained ``Economic'' relevant topics and ``Non-Economic'' relevant topics.} The treatment distribution mean is more negative, implying individuals in the treatment shifted towards the truth across all of these conditions, and the result is significant.}
    \label{fig:finding3_split_econ}
\end{figure}

%\subsection*{Self-Reported Belief Updates}

%\subsection*{Network-Level Following Trends}

% \begin{figure}
%     \centering
%     \includegraphics[width=1\linewidth]{supplemental_imgs/assortativity_control.png}
%     \caption{Assortativity Measures in Control, Comparing Stage 2 and Stage 3.}
%     \label{fig:assortativity_control}
% \end{figure}

% \begin{figure}
%     \centering
%     \includegraphics[width=1\linewidth]{supplemental_imgs/assortativity_case.png}
%     \caption{Assortativity Measures in Case, Comparing Stage 2 and Stage 3.}
%     \label{fig:assortativity_case}
% \end{figure}

%%%%%%%%%%%%%%%%%%%%%%%%%%%misinfo score diff demography 
%\subsection*{Change in misinformation score based on demography}

%%%%%%%%%%%%%%%%%%%%

%%%%%%%%%%%%%%%% SUPPLEMENTARY TABLES %%%%%%%%%%%%%%%

% \begin{table} % Do not use \begin{table*}
% 	\centering
% 	% Captions go above tables
% 	\caption{\textbf{All captions must start with a short bold sentence, acting as a title.}
% 		Follow the assortativityme style as main text tables.
% 		If the design is similar to previous tables, avoid repetition by refering back to them.}
% 	\label{tab:sup_example} % give each table a logical label name

% 	\begin{tabular}{lccr} % four columns, alignment for each
% 		\\
% 		\hline
% 		A & B & C & D\\
% 		\hline
% 		1 & 2 & 3 & 4\\
% 		2 & 4 & 6 & 8\\
% 		3 & 5 & 7 & 9\\
% 		\hline
% 	\end{tabular}
% \end{table}

\clearpage 
\newpage

\section{Tables}

\begin{table}[h]
    \centering
    \caption{\textbf{Statements used throughout the experiments.} Part 1 of 3.}
    \label{statements_part1}
    \begin{tabular}{|p{0.33\linewidth} | p{0.12\linewidth}| p{0.25\linewidth}| p{0.25\linewidth}|}
       \hline
       \textbf{Statement} & \textbf{Factuality} & \textbf{Topic} & \textbf{Evidence} \\
       \hline
       The United States does not produce the most oil in the world. & False & Fuel and Oil, Economic, Low Salience & \href{https://www.eia.gov/tools/faqs/faq.php?id=709&t=6}{https://www.eia.gov\newline/tools/faqs/} \\
       \hline
       Electric vehicles and hybrid vehicles have surged in popularity in the United States, making up 30\% of all cars on the road. & False & Fuel and Oil, Economic, Low Salience & \href{https://www.eia.gov/todayinenergy/detail.php?id=61344}{https://www.eia.gov/\newline todayinenergy/} \\
       \hline
       While gas prices are expensive in certain states, it is consistently most expensive in California and West Coast states. & True & Fuel and Oil, Economic, Low Salience & \href{https://www.eia.gov/petroleum/gasdiesel/}{https://www.eia.gov/ \newline petroleum/gasdiesel/} \\
       \hline
       In 2023, only 10 states saw a population increase. & False & Overpopulation, Non-Economic, Low Salience & \href{https://www.census.gov/newsroom/press-releases/2023/population-trends-return-to-pre-pandemic-norms.html}{https://www.census.gov/ \newline newsroom/press-releases/2023/ \newline population-trends-return-to-pre-pandemic-norms} \\
       \hline
    \end{tabular}
    \vspace{-1.5em}
\end{table}

\begin{table}[ht]
    \centering
    \caption{\textbf{Statements used throughout the experiments.} Part 2 of 3.}
    \label{statements_part2}
    \begin{tabular}{|p{0.33\linewidth} | p{0.12\linewidth}| p{0.25\linewidth}| p{0.25\linewidth}|}
       \hline
       \textbf{Statement} & \textbf{Factuality} & \textbf{Topic} & \textbf{Evidence} \\
       \hline
              United States population trends have never recovered since COVID in 2020. & False & Overpopulation, Non-Economic, Low Salience & \href{https://www.census.gov/newsroom/press-releases/2023/population-trends-return-to-pre-pandemic-norms.html}{https://www.census.gov/ \newline newsroom/press-releases/2023/ \newline population-trends-return-to-pre-pandemic-norms}\\
       \hline
       The provisional number of births for the United States in 2023 was 3,591,328, down 2\% from 2022. & True & Overpopulation, Non-Economic, Low Salience & \href{https://www.cdc.gov/nchs/data/nvsr/nvsr73/nvsr73-02.pdf}{https://www.cdc.gov/ \newline nchs/data/nvsr/nvsr73/} \\
       \hline
       Only Americans under 50 are affected by the increased recent inflation. & False & Inflation, Economic, High Salience & \href{https://news.gallup.com/poll/644690/americans-continue-name-inflation-top-financial-problem.aspx}{https://news.gallup.com/ \newline poll/americans-continue-name-inflation-top-financial-problem} \\
       \hline
       Prices for used cars and trucks, and new vehicles have fallen since 2023. & True & Inflation, Economic, High Salience & \href{https://data.bls.gov/timeseries/CUUR0000SETA02?output_view=data}{https://data.bls.gov/ \newline timeseries/} \\
       \hline
    \end{tabular}
    \vspace{-1.5em}
\end{table}

\begin{table}[ht]
    \centering
    \caption{\textbf{Statements used throughout the experiments.} Part 3 of 3.}
    \label{statements_part3}
    \begin{tabular}{|p{0.33\linewidth} | p{0.12\linewidth}| p{0.25\linewidth}| p{0.25\linewidth}|}
       \hline
       \textbf{Statement} & \textbf{Factuality} & \textbf{Topic} & \textbf{Evidence} \\
       \hline
       The average price for airline tickets increased in August 2024 when compared to August 2023. & False & Inflation, Economic, High Salience & \href{https://data.bls.gov/timeseries/CUUR0000SETG01?output_view=data}{https://data.bls.gov/ \newline timeseries/} \\
       \hline
       The number of people in the U.S. illegally is upwards of 20, 25, maybe 30 million. & False & Immigration, Non-Economic, High Salience & \href{https://www.politifact.com/factchecks/2024/jun/11/marco-rubio/there-arent-20-million-to-30-million-immigrants-in/}{https://www.politifact \newline .com/factchecks/2024/ \newline jun/11/marco-rubio/there-arent-20-million-to-30-million-immigrants-in/} \\
       \hline
       In January 2023, the U.S. started accepting people monthly from Cuba, Haiti, Nicaragua and Venezuela under a humanitarian parole program. & True & Immigration, Non-Economic, High Salience & \href{https://www.politifact.com/factchecks/2024/may/09/rick-scott/biden-isnt-flying-migrants-to-florida-people-pay-f}{https://www.politifact \newline .com/factchecks/2024/ \newline may/09/rick-scott/biden-isnt-flying-migrants-to-florida/} \\
       \hline
       New York City has been accommodating illegal immigrants in luxury hotels. & False & Immigration, Non-Economic, High Salience & \href{https://www.factcheck.org/2024/06/factchecking-the-biden-trump-debate}{https://www.factcheck \newline.org/2024/06/ \newline factchecking-the-biden-trump-debate} \\
       \hline
    \end{tabular}
    \vspace{-1.5em}
\end{table}

%%%%%%%%%%%%%%%%%%%%%%%%%%%%%%%%%%%
\begin{table}[ht]
\caption{\textbf{Number of Articles for Each Source Utilized}. Sources were identified based on balancing political leanings and availability of full-length articles.}
\label{tablearticles}
\begin{tabular}{|p{0.3\linewidth} | p{0.4\linewidth}| p{0.25\linewidth}|}
\hline
\textbf{News Source}  & \textbf{Total}       & \textbf{Political Leaning}                                                       \\ 
\hline
Washington Examiner   & 5916 & Right      \\ \hline
New York Post         & 7821     & Right      \\ \hline
Fox News              & 7506  & Right       \\ \hline
AP News & 7574      & Center  \\ \hline
The Economist     & 7432     & Center   \\ \hline
BBC     & 7598      & Center       \\ \hline
New York Times & 7695     & Left  \\ \hline
CNN   & 7464      & Left \\ \hline
Washington Post & 7494  & Left \\ \hline
MSNBC & 7632 & Left  \\ \hline 

\textbf{All Articles} & \textbf{72594} (1715 duplicates removed) & \textbf{3 right, 3 center, 4 left} \\
\hline
\end{tabular}
\end{table}

\begin{table}[]
    \caption{\textbf{Samples of Personalized LLM Final Outputs and Input Statements.} Part 1 of 4.}
    \centering
    \begin{tabular}{|p{0.5\linewidth} |p{0.5\linewidth} |}
    \hline
       \textbf{Input Statement}  & \textbf{LLM Final Output} \\ \hline
       While gas prices are expensive in certain states, it is consistently most expensive in California and West Coast states.  &  I can confidently say that the information regarding gas prices is accurate, especially considering the credible sources like the Washington Examiner. Their reports consistently highlight that California has the highest gas prices in the nation, with figures like \$5.172 per gallon and fluctuations around \$4.96, which are significantly higher than other states. Additionally, the comparison with other West Coast states like Hawaii and Washington further supports this claim. It's essential to rely on reputable news outlets that provide consistent data to understand the economic landscape accurately. \\ \hline
        While gas prices are expensive in certain states, it is consistently most expensive in California and West Coast states.  & I can assure you that the information regarding gas prices in California and the West Coast is credible and well-supported. The New York Post has reported on the California Air Resources Board's projections, indicating that gas prices could rise significantly due to emissions reduction programs, with increases expected to reach \$1.15 per gallon by 2046. Furthermore, California's commitment to phasing out gas-powered cars and transitioning to electric vehicles by 2035 demonstrates a clear long-term strategy that will likely keep gas prices high. These facts highlight the ongoing challenges residents face with fuel costs in these regions. \\  \hline
        \end{tabular}
    \label{tab:llmsamples1}
\end{table}

\begin{table}[]
    \caption{\textbf{Samples of Personalized LLM Final Outputs and Input Statements.} Part 2 of 4.}
    \centering
    \begin{tabular}{|p{0.5\linewidth} |p{0.5\linewidth} |}
    \hline
       \textbf{Input Statement}  & \textbf{LLM Final Output} \\ \hline
        Only Americans under 50 are affected by the increased recent inflation. &     The claim that only Americans under 50 are affected by recent inflation is simply not true. As highlighted by CNN, inflation impacts all age groups, influencing essential costs like housing, rent, and services that everyone relies on, regardless of age. The economic pressures are felt widely, meaning that older Americans are just as affected by rising prices as younger ones. It's crucial to recognize that the financial challenges posed by inflation are a shared experience across all demographics, not limited to a specific age group. \\ \hline
        Only Americans under 50 are affected by the increased recent inflation. &  I understand how easy it is to believe that inflation only affects a specific group, but the reality is much broader. The Economist clearly points out that inflation is a global issue impacting people of all ages and nationalities, not just Americans under 50. It’s important to recognize that this economic challenge touches everyone, regardless of age, and we should be empathetic to the struggles faced by individuals across the world. Let's focus on the collective experience rather than narrowing it down to a single demographic. \\ \hline
        \end{tabular}
    \label{tab:llmsamples2}
\end{table}

\begin{table}[]
    \caption{\textbf{Samples of Personalized LLM Final Outputs and Input Statements.} Part 3 of 4.}
    \centering
    \begin{tabular}{|p{0.5\linewidth} |p{0.5\linewidth} |}
    \hline
       \textbf{Input Statement}  & \textbf{LLM Final Output} \\ \hline
       United States population trends have never recovered since COVID in 2020.  &     The claim that the United States population trends have never recovered since COVID-19 is misleading. According to AP News, the World Health Organization has reported a downward trend in the pandemic, with increasing population immunity due to vaccinations and infections, which has allowed the U.S. to return to pre-COVID life. Additionally, the ending of the public health emergency indicates a significant recovery in population trends since 2020. Therefore, the evidence clearly shows that the situation has improved, contradicting the assertion that recovery has not occurred. \\ \hline
        United States population trends have never recovered since COVID in 2020.  &  I believe the claim that only 10 states saw a population increase in 2023 is misleading. According to the U.S. Census Bureau, while overall population growth in the U.S. has slowed, it is essential to recognize that this is part of a broader trend influenced by declining fertility rates, not just a simple increase or decrease in specific states. The dynamics of population change are much more intricate, and many states may still be experiencing growth due to factors like migration and economic opportunities. Therefore, the assertion oversimplifies a complex issue and does not accurately reflect the realities of population trends across the country. \\  \hline
        \end{tabular}
    \label{tab:llmsamples3}
\end{table}

\begin{table}[]
    \caption{\textbf{Samples of Personalized LLM Final Outputs and Input Statements.} Part 4 of 4.}
    \centering
    \begin{tabular}{|p{0.5\linewidth} |p{0.5\linewidth} |}
    \hline
       \textbf{Input Statement}  & \textbf{LLM Final Output} \\ \hline
       In January 2023, the U.S. started accepting people monthly from Cuba, Haiti, Nicaragua and Venezuela under a humanitarian parole program.   &    I can confidently affirm that the information about the U.S. accepting individuals from Cuba, Haiti, Nicaragua, and Venezuela under a humanitarian parole program in January 2023 is accurate. This was reported by AP News, a reputable and widely respected news organization known for its commitment to factual reporting. Furthermore, the context provided by the U.S. government's support for Venezuelan-led negotiations underscores the seriousness of their humanitarian efforts. Such credible sources reinforce the validity of this information and highlight the importance of addressing migration issues in a compassionate manner. \\ \hline
        In January 2023, the U.S. started accepting people monthly from Cuba, Haiti, Nicaragua and Venezuela under a humanitarian parole program.   &  I understand that the topic of migration can be complex and sometimes overwhelming, but it's important to recognize the humanitarian efforts being made. In January 2023, the U.S. initiated a program to accept individuals from Cuba, Haiti, Nicaragua, and Venezuela, showing a commitment to support those in dire situations. Additionally, the UN's plan to distribute \$1.6 billion in aid to these migrants highlights the global recognition of their struggles and the need for assistance. By embracing this program, we are not only offering hope to those in need but also reinforcing our values of compassion and support for vulnerable populations.  \\  \hline
        
        % In January 2023, the U.S. started accepting people monthly from Cuba, Haiti, Nicaragua and Venezuela under a humanitarian parole program.  & 
        %  I can assure you that the information regarding the U.S. humanitarian parole program is credible and well-supported by reputable sources. CNN reported that the U.S. State Department confirmed the release of 222 Nicaraguan political prisoners who were granted humanitarian parole, allowing them to stay in the U.S. for two years and apply for asylum. This aligns with the program initiated in January 2023, which also includes individuals from Cuba, Haiti, and Venezuela. The backing of such significant news outlets reinforces the validity of this information.  \\ \hline
        \end{tabular}
    \label{tab:llmsamples4}
\end{table}

\begin{table}[t]
\centering
\caption{\textbf{Participant counts across demographic groups for Control and Treatment conditions.} We see relatively similar representation between the control and the treatment.}
\label{tab:demographics}
\begin{tabular}{|l|l|c|c|}
\hline
\textbf{Category} & \textbf{Group} & \textbf{Control} & \textbf{Treatment} \\
\hline
\multirow{6}{*}{Age Range} 
  & 18--24 & 85 & 72 \\
  & 25--34 & 235 & 229 \\
  & 35--44 & 186 & 185 \\
  & 45--54 & 96 & 105 \\
  & 55--64 & 45 & 52 \\
  & 65+    & 20 & 14 \\
\hline
\multirow{4}{*}{Gender}
  & Male                   & 271 & 262 \\
  & Female                 & 374 & 375 \\
  & Non-Binary             & 19  & 18 \\
  % & Prefer not to answer   & 3   & 2  \\
\hline
\multirow{6}{*}{Race}
  & White                                  & 419 & 440 \\
  & Black or African American              & 122 & 110 \\
  & Hispanic                               & 53  & 49  \\
  & Asian                                  & 43  & 39  \\
  & American Indian or Alaska Native       & 1   & 4   \\
  & Other                                  & 29  & 15  \\
\hline
\multirow{5}{*}{Education Level}
  & High School Diploma            & 187 & 211 \\
  & Bachelor's Degree              & 252 & 233 \\
  & Associate Degree               & 107 & 65  \\
  & Graduate or Professional Degree & 112 & 136 \\
  & Other                          & 9   & 12  \\
\hline
\multirow{4}{*}{Political}
  & Democrat   & 314 & 306 \\
  & Republican & 126 & 139 \\
  & Independent & 207 & 190 \\
  & Other       & 20  & 22  \\
\hline
\multirow{5}{*}{Political Strength}
  & Very Strong     & 203 & 215 \\
  & Somewhat Strong & 218 & 249 \\
  & Neutral         & 184 & 154 \\
  & Somewhat Weak   & 45  & 34  \\
  & Very Weak       & 17  & 5   \\
\hline
\end{tabular}
\label{tab:participant_dem}
\end{table}

\begin{table}[ht]
\caption{\textbf{Chi‑square test statistics and p‑values for each demographic feature.} These findings suggests that the distribution of control and treatment is similar at $ \alpha = 0.05$ for age range, gender, race and politics. The distribution for education level and political strength is different.}
\centering
\begin{tabular}{|l|r|r|}
\hline
\textbf{Feature}            & \textbf{\(\chi^2\)} & \textbf{P-value} \\ 
\hline
Age Range                   &  3.05 & 0.6925 \\ 
\hline
Gender                      &  0.30 & 0.9591 \\ 
\hline
Race                        &  7.67 & 0.1757 \\ 
\hline
Education Level             & 15.12 & 0.0045 \\ 
\hline
Political                   &  1.49 & 0.6849 \\ 
\hline
Political Strength          & 13.07 & 0.0110 \\ 
\hline
\end{tabular}
\label{tab:chi2_demographics}
\end{table}

\begin{table}[htbp]
  \caption{\textbf{Mann--Whitney U Test Results Comparing  Distributions of Personality Traits for treatment and control.} Our results suggest that personality distributions are similar at $\alpha = 0.05$.}
  \centering
  % increase row height
  \renewcommand{\arraystretch}{1.2}
  % increase horizontal padding
  \setlength{\tabcolsep}{8pt}
  \begin{tabular}{|l|c|c|}
    \hline
    \textbf{Trait} & \textbf{Mann-Whitney U Statistic} & \textbf{P-value} \\
    \hline
    Extroversion      & 226158.0 & 0.3063 \\
    Agreeableness     & 215744.5 & 0.6241 \\
    Conscientiousness & 209843.5 & 0.1763 \\
    Openness          & 224907.5 & 0.3972 \\
    Neuroticism       & 222953.5 & 0.5774 \\
    \hline
  \end{tabular}
  \label{tab:mwu_personality}
\end{table}

\begin{table}[h]
\caption{\textbf{Mean and Standard Deviation for Control Pre-Election and Post-Election Conditions}. Results that show lack of significance in Table \ref{tab:controlproofofinsig}.}
\label{tab:controlpvp}
\begin{tabular}{|l|l|l|}
\hline
\textbf{Control Condition}        & \textbf{Mean}   & \textbf{Standard Deviation} \\ \hline
Pre-Election, All Conditions                    & -0.114 & 0.42               \\ \hline
Post-Election, All Conditions                   & -0.91  & 0.484              \\ \hline
Pre-Election, True Statements                   & -0.117 & 0.623              \\ \hline
Post-Election, True Statements                  & -0.101 & 0.494              \\ \hline
Pre-Election, False Statements                  & -0.01  & 0.648              \\ \hline
Post-Election, False Statements                 & -0.087 & 0.637              \\ \hline
Pre-Election, High Salience                     & -0.093 & 0.394              \\ \hline
Post-Election, High Salience                    & -0.033 & 0.406              \\ \hline
Pre-Election, Low Salience                      & -0.134 & 0.443              \\ \hline
Post-Election, Low Salience                     & -0.15  & 0.546              \\ \hline
Pre-Election, Economic Relevance                & -0.143 & 0.484              \\ \hline
Post-Election, Economic Relevance               & -0.094 & 0.509              \\ \hline
Pre-Election, No Economic Relevance             & -0.083 & 0.338              \\ \hline
Post-Election, No Economic Relevance            & -0.088 & 0.458              \\ \hline
Pre-Election, Trust in Some Political News      & -0.173 & 0.53               \\ \hline
Post-Election, Trust in Some Political News     & -0.173 & 0.641              \\ \hline
Pre-Election, Skeptical of Some Political News  & -0.105 & 0.429              \\ \hline
Post-Election, Skeptical of Some Political News & -0.085 & 0.412              \\ \hline
Pre-Election, Skeptical of All Political News   & -0.091 & 0.0367             \\ \hline
Post-Election, Skeptical of All Political News  & -0.102 & 0.462              \\ \hline
Pre-Election, Average Follow Signal, Minimum Removed & 2.031 & 0.49 \\ \hline
Post-Election, Average Follow Signal, Minimum Removed & 2.0 & 0.474 \\ \hline
\end{tabular}
\end{table}

\begin{table}[h]
\caption{\textbf{Mean and Standard Deviation for Treatment Pre-Election and Post-Election Conditions}. Results that show lack of significance in Table \ref{tab:caseproofofinsig}.}
\label{tab:casepvp}
\begin{tabular}{|l|l|l|}
\hline
\textbf{Treatment Condition}                         & \textbf{Mean} & \textbf{Standard Deviation} \\ \hline
Pre-Election, All Conditions                    & -0.341        & 0.538                       \\ \hline
Post-Election, All Conditions                   & -0.367        & 0.495                       \\ \hline
Pre-Election, True Statements                   & -0.379        & 0.823                       \\ \hline
Post Election, True Statements                  & -0.305        & 0.716                       \\ \hline
Pre-Election, False Statements                  & -0.329        & 0.69                        \\ \hline
Post Election, False Statements                 & -402          & 0.641                       \\ \hline
Pre-Election, High Salience                     & -0.288        & 0.475                       \\ \hline
Post Election, High Salience                    & -0.294        & 0.427                       \\ \hline
Pre-Election, Low Salience                      & -0.394        & 0.589                       \\ \hline
Post Election, Low Salience                     & -0.439        & 0.545                       \\ \hline
Pre-Election, Economic Relevance                & -0.33         & 0.58                        \\ \hline
Post-Election, Economic Relevance               & -0.358        & 0.523                       \\ \hline
Pre-Election, No Economic Relevance             & -0.355        & 0.485                       \\ \hline
Post-Election, No Economic Relevance            & -0.377        & 0.465                       \\ \hline
Pre-Election, Trust in Some Political News      & -0.306        & 0.527                       \\ \hline
Post-Election, Trust in Some Political News     & -0.325        & 0.466                       \\ \hline
Pre-Election, Skeptical of Some Political News  & -0.38         & 0.499                       \\ \hline
Post-Election, Skeptical of Some Political News & -0.421        & 0.545                       \\ \hline
Pre-Election, Skeptical of All Political News   & -0.354        & 0.646                       \\ \hline
Post-Election, Skeptical of All Political News  & -0.314        & 0.434                       \\ \hline
Pre-Election, Average Follow Signal, Minimum Removed & 1.536 & 0.641 \\ \hline
Post-Election, Average Follow Signal, Minimum Removed & 1.572 & 0.585 \\ \hline
\end{tabular}
\end{table}

\begin{table}[h]
\caption{\textbf{Determined Variance and P Value for Control Pre-Election v. Post-Election Density Plot Distribution Comparisons}. Results show no significant difference across all measures between Pre-Election and Post-Election.}
\label{tab:controlproofofinsig}
\begin{tabular}{|p{0.4\linewidth} | p{0.4\linewidth}| p{0.2\linewidth}|}

\hline
\textbf{Control Condition \newline  
 (Pre-Election v. Post-Election)} & \textbf{Determined Variance \newline (Equal or Unequal)} & \textbf{P Value} \\ 
\hline
All Conditions                                    & Equal                                           & 0.533            \\ \hline
True Statements Only                              & Equal                                           & 0.74             \\ \hline
False Statements Only                             & Unequal                                         & 0.758            \\ \hline
High Salience Only                                & Equal                                           & 0.179            \\ \hline
Low Salience Only                                 & Equal                                         & 0.762            \\ \hline
Economic Relevance Only                           & Equal                                           & 0.376            \\ \hline
Non-Economic Relevance Only                       & Equal                                           & 0.91             \\ \hline
Trust in Some Political News Only                 & Equal                                           & 0.999            \\ \hline
Skeptical of Some Political News Only             & Equal                                           & 0.675            \\ \hline
Skeptical of All Political News Only              & Equal                                           & 0.86             \\ \hline
Average Follow Signal, Minimum Removed & Equal & 0.452 \\ \hline
\end{tabular}
\end{table}

\begin{table}[h]
\caption{\textbf{Determined Variance and P Value for Treatment Pre-Election v. Post-Election Density Plot Distribution Comparisons}. Results show no significant difference between Pre-Election and Post-Election across all measures.}
\label{tab:caseproofofinsig}
\begin{tabular}{|p{0.45\linewidth} | p{0.3\linewidth}| p{0.2\linewidth}|}
\hline
\textbf{Treatment Condition (Pre-Election v. Post-Election)} & \textbf{Determined Variance (Equal or Unequal)} & \textbf{P Value} \\ \hline
All Conditions & Equal & 0.523 \\ \hline
True Statements Only & Equal & 0.227 \\ \hline
False Statements Only & Equal & 0.158 \\ \hline
High Salience Only & Equal & 0.898 \\ \hline
Low Salience Only & Equal & 0.469 \\ \hline
Economic Relevance Only & Equal & 0.63 \\ \hline
Non-Economic Relevance Only & Equal & 0.695 \\ \hline
Trust in Some Political News Only & Equal & 0.831 \\ \hline
Skeptical of Some Political News Only & Equal & 0.49 \\ \hline
Skeptical of All Political News Only & Unequal & 0.613 \\ \hline
Average Follow Signal, Minimum Removed & Equal & 0.458 \\ \hline
\end{tabular}
\end{table}

\begin{table}[h]
\caption{\textbf{Absolute Difference and P Value for Control Pre-Election v. Post-Election Proportion Comparisons}. Results show no significant difference between Pre-Election and Post-Election across all measures.}
\label{tab:controlproofpropinsig}
\begin{tabular}{|p{0.4\linewidth} | p{0.4\linewidth}| p{0.2\linewidth}|}
\hline
\textbf{Control Condition (Pre-Election v. Post-Election)} & \textbf{Absolute Value of Proportion Difference} & \textbf{P Value} \\ \hline
More Informed 'Yes' & 0.01 & 0.931 \\ \hline
More Informed 'No' & 0.01 & 0.731 \\ \hline
More Informed 'Maybe' & 0.02 & 0.602 \\ \hline
Changed Mind 'Yes' & 0 & 0.969 \\ \hline
Changed Mind 'No' & 0.02 & 0.668 \\ \hline
Changed Mind 'Maybe' & 0.02 & 0.603 \\ \hline
Bot Involved 'Yes' & 0.01 & 0.774 \\ \hline
Bot Involved 'No' & 0 & 0.888 \\ \hline
Bot Involved 'Maybe' & 0.01 & 0.918 \\ \hline
\end{tabular}
\end{table}

\begin{table}[h]
\caption{\textbf{Absolute Difference and P Value for Treatment Pre-Election v. Post-Election Proportion Comparisons}. Results show no significant difference between Pre-Election and Post-Election across all measures.}
\label{tab:caseproofpropinsig}
\begin{tabular}{|p{0.4\linewidth} | p{0.4\linewidth}| p{0.2\linewidth}|}
\hline
\textbf{Treatment Condition (Pre-Election v. Post-Election)} & \textbf{Absolute Value of Proportion Difference} & \textbf{P Value} \\ \hline
More Informed `Yes' & 0 & 0.949 \\ \hline
More Informed `No' & 0.05 & 0.203 \\ \hline
More Informed `Maybe' & 0.04 & 0.108 \\ \hline
Changed Mind `Yes' & 0.02 & 0.556 \\ \hline
Changed Mind `No' & 0.01 & 0.815 \\ \hline
Changed Mind `Maybe' & 0.04 & 0.316 \\ \hline
Bot Involved `Yes' & 0 & 0.955 \\ \hline
Bot Involved `No' & 0.01 & 0.782 \\ \hline
Bot Involved `Maybe' & 0.01 & 0.719 \\ \hline
0 Rounds Followed Bot & 0.01 & 0.681 \\ \hline
1 Round Followed Bot & 0.02 & 0.464 \\ \hline
2 Rounds Followed Bot & 0.02 & 0.545 \\ \hline
3 Rounds Followed Bot & 0.01 & 0.698 \\ \hline
\end{tabular}
\end{table}

%#############ROBUSTNESS#############%
\begin{table}[h]
    \centering   
    \caption{\textbf{Mean, Standard Deviation, and Confidence Intervals for Shift Results for Robustness Measures}. All confidence intervals between control and treatment for a given condition are non-overlapping, implying significance.}
    \label{tab:robustness}
    \begin{tabular}{|p{0.4\linewidth} | p{0.15\linewidth}| p{0.15\linewidth}| p{0.3\linewidth}|}
    \hline
         \textbf{Condition} & \textbf{Mean} & \textbf{Standard Deviation} & \textbf{Confidence Interval}  \\\hline
         Control, True Condition & -0.108 & 0.636 & (-0.158, -0.058)\\\hline
         Treatment, True Condition & -0.341 & 0.771 & (-0.401, -0.281) \\\hline
         Control, False Condition & -0.094 & 0.571 & (-0.139, -0.05)\\\hline
         Treatment, False Condition & -0.366 & 0.667 & (-0.417, -0.315) \\\hline
         Control, High Salience Condition & -0.062 & 0.401 & (-0.106, -0.018) \\\hline
         Treatment, High Salience Condition & -0.291 & 0.451 & (-0.341, -0.242) \\\hline
         Control, Low Salience Condition & -0.142 & 0.497 & (-0.196, -0.088)\\\hline
         Treatment, Low Salience Condition & -0.417 & 0.568 & (-0.478, -0.355) \\\hline
         Control, Economic Condition & -0.118 & 0.498 & (-0.172, -0.064) \\\hline
         Treatment, Economic Condition & -0.344 & 0.553 & (-0.402, -0.285) \\\hline
         Control, Non-Economic Condition & -0.086 & 0.404 & (-0.131, -0.041)\\\hline
         Treatment, Non-Economic Condition & -0.366 & 0.553 & (-0.419, -0.314) \\\hline
        Control, Skeptical of All Political News Condition & -0.096 & 0.417 & (-0.156, -0.037)\\\hline
         Treatment, Skeptical of All Political News Condition & -0.334 & 0.552 & (-0.414, -0.255) \\\hline
        Control, Skeptical of Some Political News Condition & -0.095 & 0.421 & (-0.144, -0.046)\\\hline
         Treatment, Skeptical of Some Political News Condition & -0.401 & 0.523 & (-0.46, -0.342) \\\hline
         Control, Trust Some Political News Condition & -0.173 & 0.587 & (-0.291, -0.056) \\\hline
        Treatment, Trust Some Political News Condition & -0.316 & 0.495 & (-0.407, -0.226) \\\hline
    \end{tabular}
\end{table}

% \begin{table}[h]
%      \centering
%     \caption{\textbf{T Test Results from Overlapping Confidence Intervals, Main Figures}. Results indicate that while distributions have overlapping confidence intervals, the findings are still significant. Only the follow signal had overlap in the confidence intervals.}
%     \label{tab:ttestfollow}
%     \begin{tabular}{|l|l|l|l|}
%     \hline
%          Condition 1 &  Condition 2 & Type of Test & p value\\\hline
%          Control Follow & Case Follow & Equal Variance & 0.025 \\\hline
%     \end{tabular}
% \end{table}

\begin{table}[h]
    \centering
    \caption{\textbf{T Test Results from Overlapping Confidence Intervals, Robustness Measures}. Only the ``Trust Some'' result had overlap, and the finding is insignificant.}
    \label{tab:ttesttrustsome}
    \begin{tabular}{|l|l|l|l|}
    \hline
         \textbf{Condition 1} &  \textbf{Condition 2} & \textbf{Type of Test} & \textbf{P Value}\\\hline
         Control Trust Some & Treatment Trust Some & Equal Variance & 0.055 \\\hline
    \end{tabular}
\end{table}

\begin{table}[h]
    \centering
    \caption{\textbf{Proportion Tests for Main Paper Findings}. Results from comparing proportions for ``Changed Mind'', ``More Informed'', and ``Bot Involved'' exit survey questions, which are all significant except ``Changed Mind `Maybe''' and ``More Informed `Maybe'''.}
    \label{tab:proportiontestsmain}
    \begin{tabular}{|l|l|l|l|}
    \hline
         \textbf{Condition} & \textbf{Control Proportion} & \textbf{Treatment Proportion} & \textbf{P Value}\\\hline
         Changed Mind `Yes' & 0.24 & 0.3 & 0.02 \\\hline
         Changed Mind `No' & 0.59 & 0.52 & 0.011 \\\hline
         Changed Mind `Maybe' & 0.17 & 0.18 & 0.547 \\\hline

         More Informed `Yes' & 0.3 &  0.41 & 0.0002 \\\hline
         More Informed `No' & 0.49 &  0.38 & 0.0004 \\\hline
         More Informed `Maybe' & 0.21 & 0.21 & 0.872 \\\hline

          Bot Involved `Yes' & 0.22 & 0.51 & 6.39 e-26 \\\hline
         Bot Involved `No' & 0.47 & 0.26 & 2.21 e-13\\\hline
         Bot Involved `Maybe' & 0.31 & 0.23 & 0.0005 \\\hline
    \end{tabular}
\end{table}

\begin{table}[htbp]
\centering
\caption{\textbf{Regression Results for Whether Individuals Moved Towards the Truth in Treatment.} None of the variable coefficients are significant.}
\label{tab:misinfo_diff_case_reg}
\begin{tabular}{|l|r|r|r|r|r|r|}
\hline
\textbf{Variable} & \textbf{Coef.} & \textbf{Std. Err.} & \textbf{t} & \textbf{P$>$$|$t$|$} & \textbf{[95\% Conf. Low]} & \textbf{[95\% Conf. High]} \\
\hline
female             & -0.0224 & 0.0425 & -0.53 & 0.599 & -0.1058 &  0.0610 \\ \hline
age                &  0.0361 & 0.0185 &  1.95 & 0.051 & -0.0002 &  0.0725 \\ \hline
education          &  0.0184 & 0.0181 &  1.02 & 0.309 & -0.0171 &  0.0538 \\ \hline
nonwhite           &  0.0372 & 0.0456 &  0.82 & 0.415 & -0.0524 &  0.1268 \\ \hline
democrat           & -0.0049 & 0.0418 & -0.12 & 0.906 & -0.0869 &  0.0771 \\ \hline
republican         & -0.1059 & 0.1040 & -1.02 & 0.309 & -0.3100 &  0.0983 \\ \hline
extroversion       &  0.0133 & 0.0106 &  1.26 & 0.209 & -0.0075 &  0.0341 \\ \hline
agreeableness      & -0.0136 & 0.0115 & -1.18 & 0.239 & -0.0363 &  0.0091 \\ \hline
conscientiousness  &  0.0274 & 0.0145 &  1.89 & 0.059 & -0.0010 &  0.0559 \\ \hline
neuroticism        &  0.0131 & 0.0115 &  1.14 & 0.254 & -0.0095 &  0.0357 \\ \hline
openness           & -0.0100 & 0.0101 & -0.98 & 0.326 & -0.0298 &  0.0099 \\ \hline
\_cons             & -0.6424 & 0.1849 & -3.47 & 0.001 & -1.0055 & -0.2793 \\ 
\hline
\end{tabular}
\end{table}

\begin{table}[h]
\centering
\caption{\textbf{Regression Results for Whether Individuals Moved towards the Truth in Control.} None of the variable coefficients are significant.}
\label{tab:misinfo_diff_control_reg}
\begin{tabular}{|l|r|r|r|r|r|r|}
\hline
\textbf{Variable} & \textbf{Coef.} & \textbf{Std. Err.} & \textbf{t} & \textbf{P$>$$|$t$|$} & \textbf{[95\% Conf. Low]} & \textbf{[95\% Conf. High]} \\
\hline
female             &  0.0014 & 0.0373 &  0.04 & 0.970 & -0.0717 &  0.0746 \\ \hline
age                &  0.0048 & 0.0162 &  0.30 & 0.767 & -0.0270 &  0.0366 \\ \hline
education          & -0.0053 & 0.0179 & -0.30 & 0.768 & -0.0404 &  0.0299 \\ \hline
nonwhite           & -0.0347 & 0.0382 & -0.91 & 0.364 & -0.1098 &  0.0404 \\ \hline
democrat           & -0.0219 & 0.0369 & -0.59 & 0.554 & -0.0944 &  0.0506 \\ \hline
republican         & -0.0550 & 0.1388 & -0.40 & 0.692 & -0.3275 &  0.2175 \\ \hline
extroversion       &  0.0028 & 0.0098 &  0.29 & 0.772 & -0.0165 &  0.0222 \\ \hline
agreeableness      & -0.0170 & 0.0108 & -1.57 & 0.116 & -0.0383 &  0.0042 \\ \hline
conscientiousness  &  0.0142 & 0.0123 &  1.16 & 0.248 & -0.0099 &  0.0382 \\ \hline
neuroticism        &  0.0060 & 0.0097 &  0.62 & 0.538 & -0.0130 &  0.0249 \\ \hline
openness           & -0.0058 & 0.0095 & -0.62 & 0.538 & -0.0245 &  0.0128 \\ \hline
\_cons             & -0.0656 & 0.1493 & -0.44 & 0.661 & -0.3588 &  0.2276 \\ 
\hline
\end{tabular}
\end{table}

\begin{table}[h]
\centering
\caption{Regression Results for whether individuals followed the bot}
\label{tab:multilinear_reg}
\begin{tabular}{|l|r|r|r|r|r|r|}
\hline
\textbf{Variable} & \textbf{Coef.} & \textbf{Std. Err.} & \textbf{t} & \textbf{P$>$$|$t$|$} & \textbf{[95\% Conf. Low]} & \textbf{[95\% Conf. High]} \\
\hline
female             & -0.0742 & 0.0342 & -2.17 & 0.030 & -0.1413 & -0.0070 \\ \hline
age                & -0.0468 & 0.0145 & -3.24 & 0.001 & -0.0752 & -0.0184 \\ \hline
education          & -0.0045 & 0.0142 & -0.32 & 0.751 & -0.0324 &  0.0234 \\ \hline
nonwhite           & -0.0387 & 0.0359 & -1.08 & 0.281 & -0.1091 &  0.0317 \\ \hline
democrat           &  0.0275 & 0.0334 &  0.82 & 0.411 & -0.0381 &  0.0930 \\ \hline
republican         &  0.0290 & 0.0788 &  0.37 & 0.713 & -0.1256 &  0.1837 \\ \hline
extroversion       &  0.0137 & 0.0083 &  1.66 & 0.098 & -0.0025 &  0.0300 \\ \hline
agreeableness      & -0.0015 & 0.0092 & -0.16 & 0.873 & -0.0196 &  0.0166 \\ \hline
conscientiousness  & -0.0068 & 0.0100 & -0.67 & 0.501 & -0.0265 &  0.0130 \\ \hline
neuroticism        & -0.0012 & 0.0090 & -0.14 & 0.890 & -0.0189 &  0.0164 \\ \hline
openness           &  0.0130 & 0.0088 &  1.47 & 0.142 & -0.0044 &  0.0303 \\ \hline
\_cons             &  0.5959 & 0.1380 &  4.32 & 0.000 &  0.3250 &  0.8667 \\ 
\hline
\end{tabular}
\end{table}

%################################demographic misinfo_diff #############
\begin{table}[h]
\caption{\textbf{Within-Group Analysis Results for Change in Misinformation Score for Education.} All results are insignificant.}
\centering
\begin{tabular}{|l|c|c|c|}
\hline
Group & F-Statistic & P-Value & Significant \\
\hline
Control & 0.207 & 0.934 & False \\ \hline
Treatment    & 1.648 & 0.160 & False \\ 
\hline
\end{tabular}
\label{tab:within_group_analysis_education}
\end{table}

\begin{table}[ht]
\caption{\textbf{Between-Group Analysis, Results for Change in Misinformation Score for Education.} All categories but ``other'' are significant.}
\centering
\begin{tabular}{|l|c|c|c|}
\hline
\textbf{Education Level} & \textbf{T-Statistic} & \textbf{P-Value} & \textbf{Significant} \\
\hline
Bachelor's degree                & 5.843 & 9.773e-09 & True  \\ \hline
other                            & 2.080 & 0.053      & False \\ \hline
High school diploma              & 5.884 & 8.848e-09 & True  \\ \hline
Associate degree                 & 2.307 & 0.023      & True  \\ \hline
Graduate or professional degree  & 3.082 & 0.002      & True  \\ 
\hline
\end{tabular}
\label{tab:between_group_analysis_education}
\end{table}

\begin{table}[htbp]
\caption{\textbf{Within-Group Analysis Results for Change in Misinformation Score for Gender.} All categories are insignificant.}
\centering
\begin{tabular}{|l|r|r|c|}
\hline
\textbf{Group} & \textbf{F-Statistic} & \textbf{P-Value} & \textbf{Significant} \\
\hline
Control & 9.968e-04 & 0.975 & False \\ \hline
Treatment    & 0.101     & 0.751 & False \\ 
\hline
\end{tabular}
\label{tab:within_group_analysis_gender}
\end{table}

\begin{table}[htbp]
\caption{\textbf{Between-Group Analysis, Results for Change in Misinformation Score for Gender.} All categories are significant.}
\centering
\begin{tabular}{|l|r|r|c|}
\hline
\textbf{Gender} & \textbf{T-Statistic} & \textbf{P-Value} & \textbf{Significant} \\
\hline
Male   & 5.990 & 4.023e-09 & True \\ \hline
Female & 6.803 & 2.135e-11 & True \\ 
\hline
\end{tabular}
\label{tab:between_group_analysis_gender}
\end{table}

\begin{table}[htbp]
\caption{\textbf{Within-Group Analysis Results for Change in Misinformation Score for Race.} All categories are insignificant.}
\centering
\begin{tabular}{|l|r|r|c|}
\hline
\textbf{Group} & \textbf{F-Statistic} & \textbf{P-Value} & \textbf{Significant} \\
\hline
Control & 0.603 & 0.697 & False \\ \hline
Treatment    & 0.890 & 0.488 & False \\
\hline
\end{tabular}
\label{tab:within_group_analysis_race}
\end{table}

\begin{table}[t]
\caption{\textbf{Between-Group Analysis Results for Change in Misinformation Score for Race.} All categories except ``Other'' and ``American Indian or Alaska Native'' are significant.}
\centering
\begin{tabular}{|l|r|r|c|}
\hline
\textbf{Race} & \textbf{T-Statistic} & \textbf{P-Value} & \textbf{Significant} \\
\hline
White                             & 6.912 & 9.483e-12 & True \\ \hline
Other                             & 1.061 & 0.297     & False \\ \hline
Black or African American         & 3.827 & 1.715e-04 & True \\ \hline
Asian                             & 2.600 & 0.011     & True \\ \hline
Hispanic                          & 4.969 & 3.044e-06 & True \\ \hline
American Indian or Alaska Native &  --   &  --       & False \\
\hline
\end{tabular}
\label{tab:between_group_analysis_race}
\end{table}

% \begin{table}[htbp]
% \centering
% \begin{tabular}{lrrc}
% \hline
% \textbf{Group} & \textbf{F-Statistic} & \textbf{P-Value} & \textbf{Significant} \\
% \hline
% Control & 0.755 & 0.582 & False \\
% Case    & 2.586 & 0.025 & True \\
% \hline
% \end{tabular}
% \caption{Within-Group Analysis for Change in Misinformation Score by Age Range}
% \label{tab:within_group_analysis_age_range}
% \end{table}

\begin{table}[htbp]
  \caption{\textbf{Within-Group Analysis for Change in Misinformation Score by Age Range.} The treatment category is significant.}
  \centering
  % slightly increase row height
  \renewcommand{\arraystretch}{1.2}
  % adjust column separation if you like
  \setlength{\tabcolsep}{8pt}
  \begin{tabular}{|l|r|r|c|}
    \hline
    \textbf{Group} & \textbf{F-Statistic} & \textbf{P-Value} & \textbf{Significant} \\
    \hline
    Control & 0.755 & 0.582 & False \\ \hline
    Treatment    & 2.586 & 0.025 & True \\
    \hline
  \end{tabular}
  \label{tab:within_group_analysis_age_range}
\end{table}

\begin{table}[htbp]
\caption{\textbf{Between-Group Analysis for Change in Misinformation Score by Age Range.} All categories except ``65+'' are significant.}
\centering
\begin{tabular}{|l|r|r|c|}
\hline
\textbf{Age Range} & \textbf{T-Statistic} & \textbf{P-Value} & \textbf{Significant} \\
\hline
18--24 & 5.946 & 2.311e-08 & True \\ \hline
25--34 & 5.951 & 5.429e-09 & True \\ \hline
35--44 & 3.631 & 3.235e-04 & True \\ \hline
45--54 & 2.192 & 0.030     & True \\ \hline
55--64 & 2.892 & 0.005     & True \\ \hline
65+    & 1.482 & 0.149     & False \\ 
\hline
\end{tabular}
\label{tab:between_group_analysis_age_range}
\end{table}

\begin{table}[htbp]
\caption{\textbf{Within-Group Analysis Results for Change in Misinformation Score for Political Affiliation.} All categories are insignificant.}
\centering
\begin{tabular}{|l|r|r|c|}
\hline
\textbf{Group} & \textbf{F-Statistic} & \textbf{P-Value} & \textbf{Significant} \\
\hline
Control & 0.225 & 0.879 & False \\ \hline
Treatment    & 0.920 & 0.431 & False \\
\hline
\end{tabular}
\label{tab:within_group_analysis_political}
\end{table}

\begin{table}[htbp]
\caption{\textbf{Between-Group Analysis Results for Change in Misinformation Score by Political Affiliation.} All categories except ``Other'' are significant.}
\centering
\begin{tabular}{|l|r|r|c|}
\hline
\textbf{Political Affiliation} & \textbf{T-Statistic} & \textbf{P-Value} & \textbf{Significant} \\
\hline
Independent & 5.633 & 3.506e-08 & True \\ \hline
Republican  & 3.681 & 2.830e-04 & True \\ \hline
Democrat    & 6.260 & 7.428e-10 & True \\ \hline
Other       & 1.928 & 0.062     & False \\
\hline
\end{tabular}
\label{tab:between_group_analysis_political}
\end{table}

\begin{table}[htbp]
\caption{\textbf{Within-Group Analysis for Change in Misinformation Score by Strength of Political Affiliation.} All categories are insignificant.}
\centering
\begin{tabular}{|l|r|r|c|}
\hline
\textbf{Group} & \textbf{F-Statistic} & \textbf{P-Value} & \textbf{Significant} \\
\hline
Control & 0.534 & 0.711 & False \\ \hline
Treatment    & 0.739 & 0.565 & False \\
\hline
\end{tabular}
\label{tab:within_group_analysis_pol_strength}
\end{table}

\begin{table}[htbp]
\caption{\textbf{Between-Group Analysis for Change in Misinformation Score by Strength of Political Affiliation.} All categories except ``Very Weak'' are significant.}
\centering
\begin{tabular}{|l|r|r|c|}
\hline
\textbf{Political Strength} & \textbf{T-Statistic} & \textbf{P-Value} & \textbf{Significant} \\
\hline
Very Weak       & 0.021 & 0.984     & False \\ \hline
Somewhat Weak   & 2.978 & 0.005     & True \\ \hline
Neutral         & 5.498 & 8.142e-08 & True \\ \hline
Somewhat Strong & 4.573 & 6.217e-06 & True \\ \hline
Very Strong     & 5.234 & 2.664e-07 & True \\
\hline
\end{tabular}
\label{tab:between_group_analysis_pol_strength}
\end{table}

%%%%%%%%%%%%Demographic difference related to followed or not followed%%%%%%%%%%%%%%%%%%%%%
%NOT doing it because we have the multilinear regression. I can add though if even later we feel like we want it. 

\clearpage
\end{document}